\documentclass[
aps,prd,
showpacs,twocolumn,notitlepage,
amssymb,amsmath,amsfonts,mathrsfs,
nofootinbib,superscriptaddress,
floats,floatfix
 amsmath,amssymb,
 aps,
]{revtex4-2}

\usepackage{graphicx}
\usepackage{xcolor}
\usepackage[colorlinks=true]{hyperref}
\hypersetup{citecolor=cyan,linkcolor=magenta}
\usepackage{multirow,array}

\usepackage{amsmath}
\usepackage{graphics}
\usepackage{epstopdf}
\usepackage{enumerate}

\usepackage{savesym}
\savesymbol{tablenum}
\usepackage{siunitx}
\restoresymbol{SIX}{tablenum}
\newcommand{\eeq}{\end{equation}}
\newcommand{\beqn}{\begin{eqnarray}}
\newcommand{\eeqn}{\end{eqnarray}}
\newcommand{\pa}{\partial}

\newcommand{\cR}{{\cal{R}}}

\newcommand{\cC}{{\cal{C}}}

\usepackage{siunitx}
\usepackage{graphicx}
\usepackage{dcolumn}
\usepackage{bm}

\begin{document}

\title[]{Relativistic Roche problem for stars in precessing orbits around a spinning black hole}

\author{Matteo Stockinger}

\affiliation{Max-Planck-Institut f\"ur Gravitationsphysik (Albert-Einstein-Institut), Am M\"uhlenberg 1, D-14476 Potsdam-Golm, Germany}
\email{matteo.stockinger@aei.mpg.de}

\author{Masaru Shibata}
\affiliation{Max-Planck-Institut f\"ur Gravitationsphysik (Albert-Einstein-Institut), Am M\"uhlenberg 1, D-14476 Potsdam-Golm, Germany}
\affiliation{Center for Gravitational Physics and Quantum Information, Yukawa Institute for Theoretical Physics, Kyoto University, Kyoto, 606-8502, Japan}

\begin{abstract}
Tidal disruptions of stars on the equatorial plane orbiting Kerr black holes have been widely studied. However thus far, there have been fewer studies of stars in inclined precessing orbits around a Kerr black hole. In this paper, we use the tensor virial equations to show the presence of possible resonances in these systems for typical physical parameters of black hole-neutron star binaries in close orbits or of a white dwarf/an ordinary star orbiting a supermassive black hole. This suggests the presence of a new instability before the tidal disruption limit is encountered in such systems. 
\end{abstract}

\maketitle

\section{Introduction}\label{secI}

The classical Roche problem studies the equilibrium and stability of a self-gravitating fluid in a binary system. Its main result is the presence of a critical orbital distance, the so-called Roche limit, below which there is no stable configuration, i.e., mass shedding and tidal disruption of the star are induced~\cite{Aizenman, CS}. Extensions of these results to eccentric orbits have also been the focus of some studies \cite{Nduka,NewtEllipt}. Furthermore, the concept of the Roche limit has been extended to general relativistic scenarios, such as when the binary companion is a black hole (BH)~\cite{Rel_RocheI, Rel_RocheII, Mashhoon, Sh}. The previous studies for this focused on the star in circular equatorial orbits around the BH, and to our knowledge, the only exception is~\cite{TidalNonEq}.

In this paper, we study the case of a star in a circular orbit on the non-equatorial plane around a spinning BH. To study this, as a first step, we assume that the star is incompressible and utilize the tensor virial formalism developed by Chandrasekhar~\cite{CS}. This formalism has generally been used to study the assumed ellipsoidal shape of a self-gravitating fluid and provided a number of knowledge on the equilibrium and stability of the stars qualitatively. 

After reviewing the previous results, we show the presence of a resonance when the orbit has an inclination with respect to the equatorial plane. To compute the impact of the inclination on the shape of the star, we use a perturbative expansion assuming that the inclination angle is small: We perform this expansion up to the second order in the angle of the inclined orbit $\varepsilon=\theta-\pi/2$ and show that the resulting equations take a linear form, which is usually solvable. When this is not the case, we posit a new kind of a resonance for such systems. We show that the resonances can play an important role in the evolution of BH-neutron star (NS) binaries in close orbits or of a white dwarf/an ordinary star orbiting a supermassive BH.

This paper is organized as follows: In Sec.~\ref{virEq}, we present the working assumptions and the tensor virial equations adapted to our system, by which we derive the equilibrium solution of a star on an equatorial orbit around a BH~\cite{Rel_RocheI, Rel_RocheII, Mashhoon, Sh}. In Sec.~\ref{Per}, we present our method for the perturbative expansion, which we apply up to the second order in $\varepsilon$. In Sec.~\ref{Res}, we show the presence of new resonances in these systems, using the equations derived in Sec.~\ref{Per}. Finally, we discuss the implications of our findings in Sec.~\ref{Dis}.
Throughout this paper, we use the geometrical units of $c=G=1$ where $c$ and $G$ are the speed of light and gravitational constant, respectively.  We will not use the Einstein summation convention.

\section{Framework of the system}
\label{virEq}

\subsection{System}

The purpose of this paper is to study the tidal effect on a star orbiting a Kerr BH. We assume that the star is modeled by an incompressible fluid of mass $M$ and constant density $\rho$. This star is assumed to have an ellipsoidal shape determined by its principal semi-axes $(a_1, a_2, a_3)$. We assume that it orbits a spinning BH of mass $m$ and of spin  parameter $a$. When referencing its position relative to the BH, we will use the Boyer-Lindquist coordinates in which the line element is written as
\begin{multline}
    ds^2=-\frac{\Delta}{\Sigma}(dt-a\sin^2\theta d\phi)^2+\\\frac{\sin^2\theta}{\Sigma}[(r^2+a^2)d\varphi-adt]^2+\frac{\Sigma}{\Delta}dr^2+\Sigma d\theta^2,
\end{multline} 
where $\Delta=r^2-2mr+a^2$ and $\Sigma=r^2+a^2\cos^2\theta$. In the following we assume that $M \ll m$ and do not consider the self-gravitating effect of the stars on the orbital motion around the BH. That is, we assume that the star has a geodesic orbit around the BH.


To study the effect on the stars of the tidal force from the BH, we prepare Fermi-normal coordinates on the geodesics~\cite{Ma}. The tidal force acting on the star by the BH is evaluated in the local inertial frame on the Fermi-normal coordinates. This allows us to treat the problem in the similar way to Newtonian hydrodynamics.

The geodesics on the Kerr spacetime are described by the four constants of motion,  $E$: specific energy, $L$: specific angular momentum for the BH spin direction, $K$: the Carter constant~\cite{Carter:1968rr}, and the mass of the star. The geodesic equations are written as~(e.g.,~\cite{Bardeen:1972fi})
\beqn
{dt \over d\tau}&=&{[(r^2+a^2)^2-\Delta a^2 \sin^2\theta] E - 2m r a L
  \over \Delta \Sigma},\\
\Sigma^2\left({dr \over d\tau}\right)^2&=&
\left\{E(r^2+a^2)-aL\right\}^2
-\Delta (r^2 +K) \nonumber \\
&:=&\cR(r),\label{Req}\\
\Sigma^2 \left({d\theta \over d\tau}\right)^2&=&
K-a^2\cos^2\theta-{(aE \sin^2\theta-L)^2 \over \sin^2\theta}
\label{Thetaeq}\\
{d\varphi \over d\tau}&=&{1 \over \Delta}
          \left[{2mr aE \over \Sigma}+\left(1-{2mr \over \Sigma}\right)
            {L \over \sin^2\theta}\right],
\eeqn
where $\tau$ is an affine parameter of the geodesics. In this paper, we will assume that the star has a circular orbit with a fixed value of $r=r_0$; $\cR=0=d\cR/dr=0$ at $r=r_0$. These relations give us the two relations among $E$, $L$, and $K$.

In this paper, we consider precessing orbits with respect to the equatorial plane, i.e.,  $\theta\not=\pi/2$. For a given maximum value of $\theta$ for the orbit, we then have an additional relation among $E$, $L$, and $K$, and thus, these quantities are written as a function of $r_0$ (see, e.g., \cite{OtherSpherical, Spherical}).

In the inertial frame defined above, we refer to the axes of the corresponding orthonormal basis as 1, 2, and 3. We assume that the axis 3 points in the same direction as the orbital angular momentum of the fluid or in other words, the direction of $(\pa/\pa\theta)^\mu$. Note that our 2 and 3 axes agree with the 3 and 2 axes of \cite{Ma}, and that we changed the direction of our 2-axis. Assuming that the stellar radius, $R_\mathrm{star}$, is much smaller than the orbital separation to the BH, $r_0$, one can write the tidal tensor as~\cite{Ma}
\begin{eqnarray}
    C_{11}&=&\Big(1-3\frac{ST(r^2-a^2\cos^2\theta)}{K\Sigma^2}\cos^2\Psi\Big)I_a
    \nonumber \\
    &&+6ar\cos\theta\frac{ST}{K\Sigma^2}\cos^2\Psi I_b, \\
    C_{22}&=&\Big(1-3\frac{ST(r^2-a^2\cos^2\theta)}{K\Sigma^2}\sin^2\Psi\Big)I_a\nonumber \\
    &&+6ar\cos\theta\frac{ST}{K\Sigma^2}\sin^2\Psi I_b,\\
    C_{33}&=&\Big(1+3\frac{r^2T^2-a^2\cos^2\theta S^2}{K\Sigma^2}\Big)I_a\nonumber \\&&-6ar\cos\theta\frac{ST}{K\Sigma^2}I_b,\\
    C_{12}&=&-3\Big[(a^2\cos^2\theta-r^2)I_a\nonumber \\
    &&+2ar\cos\theta I_b\Big]\frac{ST}{K\Sigma^2}\cos\Psi\sin\Psi, 
\end{eqnarray}
\begin{eqnarray}
C_{13}&=&3\Big[-ar\cos\theta(S+T)I_a\nonumber \\
    &&+(a^2\cos^2\theta S-r^2T)I_b\Big]\frac{\sqrt{ST}}{K\Sigma^2}\cos\Psi,
\\
    C_{23}&=&-3\Big[-ar\cos\theta(S+T)I_a\nonumber \\
    &&+(a^2\cos^2\theta S-r^2T)I_b\Big]\frac{\sqrt{ST}}{K\Sigma^2}\sin\Psi, 
\end{eqnarray}
where 
\begin{eqnarray}
I_a&=&\frac{mr}{\Sigma^3}(r^2-3a^2\cos^2\theta), \\
I_b&=&\frac{ma\cos\theta}{\Sigma^3}(3r^2-a^2\cos^2\theta), 
\end{eqnarray}
$S=r^2+K$, $T=K-a^2\cos^2\theta$, and $\Psi$ is a time-dependent angle which obeys the following equation~\cite{Ma}
\begin{equation}
    {d\Psi \over d\tau}=\frac{\sqrt{K}}{\Sigma}\left(\frac{E(r^2+a^2)-aL}{r^2+K}+a\frac{L-aE\sin^2\theta}{K-a^2\cos^2\theta}\right).
\end{equation}
We note that higher-order corrections in $R_\mathrm{star}/r_0$ for the tidal tensors are found in \cite{Ishii:2005xq}. One important aspect of the tidal tensor is that $C_{23}$ and $C_{13}$ are non-zero only in the presence of the BH spin and for $\theta\not=\pi/2$. This implies that for a star in precessing orbits around a Kerr BH, a qualitatively new tidal force, which is absent in Newtonian gravity, is exerted. 

The expression of the tidal tensor can be simplified by changing the frame we are working on. Indeed in a rotating frame of rotation $\vec\Omega$ along the 3-axis with magnitude $\Omega=d\Psi/d\tau$, the tidal tensor is simplified as~\cite{Ma}
\begin{eqnarray}
    \tilde C_{11}&=&\Big(1-3\frac{ST(r^2-a^2\cos^2\theta)}{K\Sigma^2}\Big)I_a\nonumber \\
    &&~+6ar\cos\theta\frac{ST}{K\Sigma^2}I_b, \\
    \tilde C_{22}&=&I_a,\\
    \tilde C_{33}&=&\Big(1+3\frac{r^2T^2-a^2\cos^2\theta S^2}{K\Sigma^2}\Big)I_a\nonumber \\
    &&~-6ar\cos\theta\frac{ST}{K\Sigma^2}I_b,\\
    \tilde C_{12}&=&0,\\ 
    \tilde C_{13}&=&3\Big[-ar\cos\theta(S+T)I_a\nonumber \\
    &&~+(a^2\cos^2\theta S-r^2T)I_b\Big]\frac{\sqrt{ST}}{K\Sigma^2},\\
    \tilde C_{23}&=&0.
\end{eqnarray} 
In the following, all the analyses will be carried out in this frame.

\subsection{Tensor virial equations}

As already stated in Sec.~\ref{secI}, we employ the tensor virial equations~\cite{1954PhRv...96.1686P,CS} to study the equilibrium state of stars orbiting a Kerr BH. The hydrodynamics equation for the fluid on the local inertial frame is written as
\begin{equation}
    \rho\frac{dU_i}{d\tau}=-\frac{\partial P}{\partial X_i}-\rho\frac{\partial\phi}{\partial X_i}-\rho \sum_{l=1}^3 C_{il}X_l,
\end{equation} 
where $X_i=(X_1, X_2, X_3)$ denote the coordinates in the local inertial frame, $U_i$ is the velocity field of the star, $P$ the pressure, and $\phi$ the Newtonian potential of the fluid, which obeys the Poisson equation as 
\begin{equation}
    \vartriangle\phi=4\pi\rho,
\end{equation}
where $\vartriangle$ denotes the Laplacian in the flat space. We assume that the self-gravity of the star is not so strong that the equation of motion is written as in the Newtonian case. We have to keep in mind that this can introduce an error of the size of $GM/c^2R_\mathrm{star}$ in considering NSs as the star. 

As seen above, it is preferable to work in a rotating frame of angular velocity vector $\vec\Omega$, as this simplifies the expression of the tidal tensor. In this case, the equation of motion is rewritten as
\begin{multline}
    \rho\frac{du_i}{d\tau}=-\frac{\partial P}{\partial x_i}-\rho\frac{\partial\phi}{\partial x_i}+\rho(\Omega^2x_i-\delta_{i3}\Omega^2x_3)\\+2\rho\sum_{l=1}^3\epsilon_{il3}u_l\Omega
    +\rho\sum_{l=1}^3\epsilon_{il3}x_l\dot{\Omega}-\rho \sum_{l=1}^3\tilde{C}_{il}x_l,
    \label{eq23}
\end{multline}
where $x_i$ and $u_i$ denote the coordinates and the velocity field of the star in the rotating frame, $\delta_{ij}$ is the Kronecker delta, 
$\epsilon_{ijk}$ is the completely antisymmetric tensor in 3 dimensions, and $\dot \Omega=d\Omega/d\tau$.

We then assume that the pressure is given by 
\begin{equation}
    P=p_c\left(1-\sum_{i=1}^3\frac{x_i^2}{a_i^2}\right),
\end{equation} 
where $p_c$ is the central pressure and $a_i$~($i=1$--3) denote the axial lengths of the star for the corresponding directions. This setting enables us to assume that the velocity depends linearly on the position as
\begin{equation}
u_i=\sum_{k=1}^3Q_{ik}x_k, 
\end{equation}
where $Q_{ik}$ is a matrix to be defined.

We now follow the method developed by Chandrasekhar to derive the second-rank tensor virial equations~\cite{CS}.
First, we multiply $x_j$ to Eq.~(\ref{eq23}) and integrate over the volume $V$ of the entire star. Then we obtain
%
\begin{eqnarray}
&&\int_V\rho\frac{du_i}{d\tau}x_jd^3x=-\int_V\frac{\partial P}{\partial x_i}x_jd^3x-\int_V\rho\frac{\partial\phi}{\partial x_i}x_jd^3x\nonumber \\
&&~~~~~~+\int_V\rho(\Omega^2x_i-\delta_{i3}\Omega^2x_3)x_jd^3x\nonumber \\
&&~~~~~~+2\Omega\sum_{l=1}^3\epsilon_{il3}\int_Vu_lx_jd^3x+\sum_{l=1}^3\epsilon_{il3}\dot{\Omega}\int_Vx_lx_jd^3x\nonumber \\
&&~~~~~~-\sum_{l=1}^3\int_V\rho \tilde{C}_{il}x_l x_jd^3x.\label{bigEq}\end{eqnarray}
To rewrite this equation, we introduce the following integral variables: 
\begin{eqnarray}
    I_{ij}&=&\int_V\rho x_ix_jd^3x,\\
    2T_{ij}&=&\int_V\rho u_iu_jd^3x=\sum_{k=1}^3\sum_{l=1}^3Q_{ik}Q_{jl}I_{kl},\\
    J_{ij}&=&\int_V\rho u_ix_jd^3x=\sum_{k=1}^3Q_{ik}I_{kj},\\
    M_{ij}&=&-\int_V\rho x_i\partial_j\phi d^3x,\\
    \Pi&=&\int_VP d^3x.
\end{eqnarray}
Then, Eq.~(\ref{bigEq}) is written as
\begin{multline}
    \frac{d}{d\tau}J_{ij}=2T_{ij}+\Pi\delta_{ij}+M_{ij}+\Omega^2I_{ij}-\delta_{i3}\Omega^2I_{3j}\\+2\Omega\sum_{l=1}^3\epsilon_{il3}J_{lj}+\sum_{l=1}^3\epsilon_{il3}\dot{\Omega}I_{lj}-\sum_{l=1}^3\tilde{C}_{il}I_{lj}.\label{TVE}
\end{multline}
From this, we can deduce the symmetrized version
\begin{eqnarray}
    \frac{1}{2}\frac{d^2}{d\tau^2}I_{ij}&=&2T_{ij}+\Pi\delta_{ij}+M_{ij}\nonumber \\
    &&+\Omega^2 \left[I_{ij}-\frac{1}{2}(\delta_{i3}I_{3j}+\delta_{j3}I_{3i})\right]\nonumber \\
    &&+\Omega\sum_{l=1}^3(\epsilon_{il3}J_{lj}+\epsilon_{jl3}J_{li})\nonumber \\
    &&+\frac{1}{2}\dot\Omega\sum_{l=1}^3(\epsilon_{il3}I_{lj}+\epsilon_{jl3} I_{li})\nonumber \\
    &&-\frac{1}{2}\sum_{l=1}^3(\tilde{C}_{il}I_{lj}+\tilde C_{jl}I_{li}), 
\label{SymEq}
\end{eqnarray}
and the antisymmetrized version of the tensor virial relation~\cite{1954PhRv...96.1686P}, 
\begin{eqnarray}
    \frac{d}{d\tau}(J_{ij}-J_{ji})&=&-\Omega^2(\delta_{i3}I_{3j}-\delta_{j3}I_{3i})\nonumber \\
    &&+2\Omega\sum_{l=1}^3(\epsilon_{il3}J_{lj}-\epsilon_{jl3}J_{li})\nonumber \\
    &&+\dot\Omega\sum_{l=1}^3(\epsilon_{il3}I_{lj}-\epsilon_{jl3}I_{li})\nonumber \\
    &&-\sum_{l=1}^3(\tilde{C}_{il}I_{lj}-\tilde C_{jl}I_{li}).
    \label{aSymEq}
\end{eqnarray}
Equation~(\ref{aSymEq}) is interpreted as the evolution equation of the angular momentum. 

Equation~(\ref{TVE}) will serve as our basis for the exploration of the influence of the tidal force on stars orbiting a Kerr BH.

\subsection{Equatorial plane case}

When the star orbits a BH on its equatorial plane, the equations become analytically solvable. Although this case was already studied in~\cite{Sh}, we will here restate the main results, as they serve as the zeroth-order solutions for the main problem in this paper.

Assuming that the star is in equilibrium, Eq.~(\ref{SymEq}) is written as
\begin{equation}
\begin{split}
    0&=2T_{ij}+\Pi\delta_{ij}+M_{ij}+\Omega^2\left[I_{ij}-\frac{1}{2}(\delta_{i3}I_{3j}+\delta_{j3}I_{3i})\right]\\
&+\Omega\sum_{l=1}^3(\epsilon_{il3}J_{lj}+\epsilon_{jl3}J_{li})-\frac{1}{2}\sum_{k=1}^3(\tilde{C}_{ik}I_{kj}+\tilde C_{jk}I_{ki}),\label{eq34}
\end{split}
\end{equation}
while Eq.~(\ref{aSymEq}) is trivially satisfied because $\dot\Omega=0$ and off-diagonal components of $\tilde C_{ij}$ vanish: In this case, the non-zero components of the tidal tensor are 
\begin{eqnarray}
    \tilde C_{11}&=&\Omega_0^2\Big(1-3\frac{\Delta_0}{P_0}\Big), \\
    \tilde C_{22}&=&\Omega_0^2,\\
    \tilde C_{33}&=&\Omega_0^2\Big(-2+3\frac{\Delta_0}{P_0}\Big),
\end{eqnarray} 
where $\Omega_0^2=m/r_0^3$~\cite{Rel_RocheII}, $\Delta_0=r_0^2-2mr_0+a^2$, and $P_0=r_0^2-3mr_0+2am^{1/2}r_0^{1/2}$.
Like in the standard Roche problem, we can assume that the principal axes of the ellipsoid coincide with the axes of our frame.
Thus, we have~\cite{CS}
\begin{eqnarray}
    I_{ij}&=&\frac{1}{5}Ma_i^2\delta_{ij},\\
    M_{ij}&=&-2\pi\rho A_i I_{ij},\\
    A_j&=&a_1a_2a_3\int_0^\infty\frac{du}{D(a_j^2+u)}, 
\end{eqnarray} 
where
\begin{equation}
    D=\sqrt{(a_1^2+u)(a_2^2+u)(a_3^2+u)}.
\end{equation}

As we assume that the internal motion is characterized by a vorticity around the 3-axis, we set
\begin{equation}
    Q=\Lambda\begin{pmatrix}
        0 & a_1a_2^{-1} & 0 \\
        -a_1^{-1}a_2 & 0 & 0 \\
        0 & 0 & 0
    \end{pmatrix},
\end{equation} 
where $\Lambda$ is the magnitude of the vorticity. The circulation of the star in the inertial frame is then written as
\beqn
C_\Gamma&=&\pi a_1 a_2 \left (2 \Omega_0-{a_1^2 + a_2^2 \over a_1 a_2} \Lambda\right)
\nonumber \\
&=&2 \pi a_1 a_2 \Omega_0 (1-f),
\eeqn
where
\begin{equation}
\label{f}
    f=\frac{a_1^2+a_2^2}{2a_1a_2}\frac{\Lambda}{\Omega_0}.
\end{equation}
In this paper, we pay attention to the cases of $f=0$ and $f=1$. 
For $f=0$, the fluid has no internal motion in the rotating frame, while in the inertial frame, the fluid's internal motion comes only from the rotation.
On the other hand, for $f=1$, the fluid has a certain vorticity in the rotating frame, while in the inertial frame, the fluid does not have any circulation. We refer to this case as irrotational Roche-Riemann ellipsoid (IRRE)~\cite{CS}. It has been shown that this velocity configuration is typically (approximately) satisfied for binary neutron stars and BH-NS binaries in the late inspiraling stage~\cite{noVisc_BNS, noVisc_BHNS}. 

In the case of equatorial orbits, Eq.~(\ref{eq34}) becomes  
\begin{eqnarray}
    0&=&\tilde\Lambda^2-2\alpha_2\tilde\Omega\tilde\Lambda-2A_1+2\frac{p_c}{\pi\rho^2 a_1^2}+3\tilde\Omega^2\frac{\Delta_0}{P_0},\\
    0&=&\alpha_2^2\tilde\Lambda^2-2\alpha_2\tilde\Omega\tilde\Lambda-2\alpha_2^2A_2+2\frac{p_c}{\pi\rho^2 a_1^2},\\
    0&=&-2\alpha_3^2A_3+2\frac{p_c}{\pi\rho^2 a_1^2}-\alpha_3^2\tilde\Omega^2\left(3\frac{\Delta_0}{P_0}-2\right),
\end{eqnarray} 
where we defined dimensionless quantities, $\alpha_2:=a_2/a_1$, $\alpha_3:=a_3/a_1$, $\tilde\Omega:=\Omega_0/(\pi\rho)^{1/2}$, and $\tilde\Lambda:=\Lambda/(\pi\rho)^{1/2}$. 

\begin{figure}[t]
\vspace{-5mm}
\includegraphics[width=0.5\textwidth]{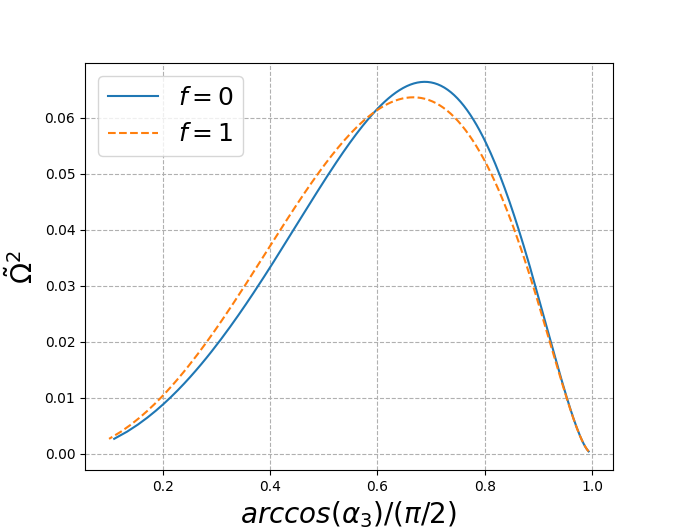}
\caption{$\tilde\Omega^2$ as a function of $\arccos(\alpha_3)/(\pi/2)$ at the ISCO for $f=0$ and $f=1$. We note that $\Delta_0/P_0=4/3$ at the ISCO irrespective of the dimensionless spin parameter, $a/m$, and hence, the curves are universal irrespective of $a/m$.}\label{0th}
\end{figure}

After fixing $f$ and eliminating $p_c/(\pi\rho^2 a_1^2)$, we can determine $\alpha_2$ and $\tilde\Omega$ as functions of $\alpha_3$. We plot the the relations between $\tilde\Omega$ and $\alpha_3$ for $f=0$ and $f=1$ in Fig.~\ref{0th}, which agree with those in \cite{Sh}. 

One notices that $\tilde\Omega$ reaches a maximum value $\tilde\Omega_\mathrm{crit}$ corresponding to $\alpha_{3,\mathrm{crit}}$. Above this value, there is no possible equilibrium configuration of the star. This gives the relativistic Roche limit, already computed in \cite{Rel_RocheI,Sh}. As in \cite{Sh}, we find that at the innermost stable circular orbit (ISCO) $\tilde\Omega_\mathrm{crit}^2\approx0.06364$ for $f=1$, and $\tilde\Omega_\mathrm{crit}^2\approx0.06640$ for $f=0$~\cite{Rel_RocheI}. As in the classical case, one can expect that the star is unstable against mass shedding/tidal disruption for $\alpha_3 <\alpha_{3, \mathrm{crit}}$ and stable for $\alpha_3>\alpha_{3, \mathrm{crit}}$. Thus, for a given value of $\rho$, $\Omega$ has a maximum value, i.e., the orbital separation $r_0$ has a minimum value, and for the smaller orbital separation, the star should start the tidal disruption process.  

\section{Stars in slightly precessing orbits}\label{Per}

\subsection{Method for the expansion}

We now explore the cases that the star does not lie on the equatorial plane. For this, we perform perturbative calculations using $\varepsilon=\theta-\pi/2$ as an infinitesimal parameter. The zeroth-order solution is determined by the equations for the equatorial orbits reviewed in the previous section. We then need to consider a small perturbation away from this configuration for $0 < \varepsilon \ll 1$. 

By perturbing the equatorial solution, we do not provide any change to the Roche limit for stars in spherical orbit as in \cite{TidalNonEq}. However this method enables us to study more in detail the dynamical behavior of the star. In particular we do not restrict the motion of the fluid to be in the 1-2 plane. While we assume such a motion in the equatorial case, we should expect a wider range of motions due to the non-diagonal elements of the tidal tensor. This intuition will be confirmed in the following analyses.

To obtain the perturbed configuration, we assume that every fluid element is displaced as $x_i \to x_i + \xi_i$ where $\xi_i$ is an infinitesimal displacement. For this, we assume the form of $\xi_i=\sum_{j=1}^3\xi_{ij}x_j$, which can satisfy the perturbation equations for the incompressible fluid self-consistently (see below). 
Furthermore, to solve this system in such an assumption, it is sufficient to consider the modified second-rank tensor virial equations~\cite{CS} together with the incompressible condition, which leads to
\begin{equation}
\label{VolConserve}
    \xi_{11}+\xi_{22}+\xi_{33}=0.
\end{equation} 

The displacement will depend on time, as the orbital value of $\theta$ does too. Thus, as a first step, we have to determine the time-dependent orbital motion. To do this we analyze the geodesic equation for $\theta$, Eq.~(\ref{Thetaeq}), which is rewritten for $\varepsilon=\theta-\pi/2 \ll 1$ as (e.g., \cite{Shibata:1994jx}) 
\begin{equation}
    r^4\left(\frac{d\varepsilon}{d\tau}\right )^2=\mathcal{C}-\varepsilon^2\left[a^2(1-E_0^2)+L_0^2\right],
\end{equation} 
where $\mathcal{C}=K-(L-aE)^2$, which vanishes for equatorial orbits, and we took into account the terms at $O(\varepsilon^2)$. Note that $E_0$ and $L_0$ denote $E$ and $L$ for equatorial circular orbits~\cite{Bardeen:1972fi} while for $\cC$ we need the second-order quantities in $\varepsilon$.
Then, for $r=r_0$, we obtain
\begin{equation}
\varepsilon(\tau)=\varepsilon_0\cos(\omega_{\theta}\tau), 
\end{equation} 
where 
\begin{eqnarray}
\omega_{\theta}&=&\frac{\sqrt{L_0^2+a^2(1-E_0^2)}}{r_0^2}\nonumber \\
&=&\sqrt{{r_0^2-4a \sqrt{mr_0}+3a^2 \over P_0}}\Omega_0, \label{omegat}\\
\varepsilon_0&=&\frac{\sqrt{\mathcal{C}}}{\omega_\theta}. 
\end{eqnarray}
Here $\varepsilon_0$ and $\sqrt{\mathcal{C}}$ are first-order parameters which we will use for the perturbative expansion. It is worthy to note at this stage that $\cos\theta(\tau)=-\varepsilon(\tau)$ at the first order.

We consider a spherical orbit with always the same radius $r_0$ but different values of  $\theta$. Following \cite{Spherical}, to study the orbit of the star, we fix $r_0$ and $\cal{C}$ and deduce the other parameters from $\cR=0=d\cR/dr$ at $r=r_0$. Thus, $E$, $L$, and $K$ depend on the inclination angle $\varepsilon_0$. These changes occur at the second order in $\varepsilon_0$ (see appendix).

The tidal tensor components can be decomposed using a Taylor expansion around $\theta=\pi/2$ : 
\begin{equation}
    \tilde C_{ij}=\sum_{k=0}^\infty \tilde C_{ij}^{(k)}\varepsilon^k.
\end{equation}
The components only contain even-order contributions, if its indices are even in the component 3, and odd-order contributions, if its indices are odd in the component 3.

Likewise, we write $\Omega$ and $\dot\Omega$ as a Taylor expansion, which contain only the terms of even power in $\varepsilon$. We note $\Omega_0=\displaystyle\sqrt{\frac{m}{r_0^3}}$ at the zeroth order of $\Omega$.

We do the same for the displacement coefficients $\xi_{ij}$. By considering the symmetry $x_3\to -x_3$, one can constrain the expression of the displacement. We thus have that if the indices of the displacement are odd in the component 3, then the series expansion only contains terms of odd power in $\varepsilon$, and likewise for indices even in the component 3, the series expansion only contains terms of even power in $\varepsilon$.

Following \cite{CS}, the integrals associated with a displacement $\xi_i$ of the fluid are given as
\begin{eqnarray}
    2\delta T_{ij}&=&\sum_{k=1}^3\biggl[Q_{jk}\frac{dV_{i;k}}{d\tau}+Q_{ik}\frac{dV_{j;k}}{d\tau}
    \nonumber \\ &&~~~~-((Q^2)_{jk}V_{i;k}+(Q^2)_{ik}V_{j;k})\biggr],\\
    \delta I_{ij}&=&V_{i;j}+V_{j;i},\\
    \delta M_{ij}&=&-2\pi\rho B_{ij}V_{ij}+\pi\rho a_i^2\delta_{ij}\sum_{l=1}^3A_{il}V_{ll},\\
    \delta\Pi&=&\int_V\delta P d^3x,
\end{eqnarray}
\begin{eqnarray}
    \delta\frac{d}{d\tau}\int_V\rho u_ix_jd^3x&=&\frac{d^2V_{i;j}}{d\tau^2}\nonumber \\
    &&+\sum_{k=1}^3 \left[Q_{ik}\frac{dV_{j;k}}{d\tau}-Q_{jk}\frac{dV_{i;k}}{d\tau}\right],~~\\
    \delta\int_V\rho u_ix_jd^3x&=&\frac{dV_{i;j}}{d\tau}+\sum_{k=1}^3\left[Q_{ik}V_{j;k}-Q_{jk}V_{i;k}\right],~~
\end{eqnarray} 
where
\begin{eqnarray}
    V_{i;j}&=&\int_V\rho \xi_i x_j d^3x=\xi_{ij}I_{jj},\\
    B_{ij}&=&a_1a_2a_3\int_0^{\infty}\frac{udu}{D(a_i^2+u)(a_j^2+u)},\\
    A_{ij}&=&a_1a_2a_3\int_0^{\infty}\frac{du}{D(a_i^2+u)(a_j^2+u)}.
\end{eqnarray} 
The resulting equations can be divided in two types as shown in the subsequent subsections. In the following, we only consider equations which contain non-zero contributions of lower-order solutions of the fluid. This allows us to focus only on the displacement induced by the inclination of the orbit. 

\subsection{First-order calculations}

To solve the first-order equations for the displacement, we select only the contribution of order $\varepsilon_0$ in the equation. If we start from Eq.~(\ref{TVE}), this gives us
\begin{widetext}
\begin{multline}
    \frac{d^2}{d\tau^2}V_{i;j}-2\sum_{k=1}^3Q_{jk}\frac{dV_{i;k}}{d\tau}=-\sum_{l=1}^3((Q^2)_{jl}V_{i;l}+(Q^2)_{il}V_{j;l})-2\pi\rho B_{ij}V_{ij}+\pi\rho a_i^2\delta_{ij}\sum_{l=1}^3 A_{il}V_{ll}\\ +\Omega_0^2V_{ij}-\delta_{i3}\Omega_0^2V_{3j}+\delta_{ij}\delta\Pi+2\sum_{l=1}^3\epsilon_{il3}\Omega_0\left(\frac{d}{d\tau}V_{l;j}+\sum_{k=1}^3(Q_{lk}V_{j;k}-Q_{jk}V_{l;k})\right)-\sum_{k=1}^3\tilde{C}^{(0)}_{ik}V_{kj}-\sum_{k=1}^3\tilde{C}^{(1)}_{ik}I_{kj},
\end{multline}
\end{widetext} with $V_{ij}=V_{i;j}+V_{j;i}$.

As before, we can write the corresponding symmetrized and antisymmetrized equations. 
We only need to consider four of these equations at first order. Indeed only the equations for the indices $(i,j)=(1,3), (3,1), (2,3), (3,2)$ contain the non-zero contribution from the inertia tensor. 

Since the first-order perturbative quantities should vary periodically in time with the angular velocity $\omega_\theta$, we use the ansatz for the variables as
\begin{eqnarray}
     V_{1;3}&=&V_{1;3}^{(1)}\cos(\omega_\theta\tau),\\
     V_{3;1}&=&V_{3;1}^{(1)}\cos(\omega_\theta\tau),\\
     V_{2;3}&=&V_{2;3}^{(1)}\sin(\omega_\theta\tau),\\
     V_{3;2}&=&V_{3;2}^{(1)}\sin(\omega_\theta\tau).
\end{eqnarray}
Then the symmetrized equations become 
\begin{eqnarray}
&&(-\omega_\theta^2+4\pi\rho B_{13}-\Omega_0^2+\tilde{C}_{11}+\tilde C_{33})V_{1;3}^{(1)}\nonumber \\
&&+(-\omega_\theta^2+4\pi\rho B_{13}-\Omega_0^2-2\Lambda^2\nonumber\\
&&+2\Omega_0\Lambda \alpha_2+\tilde{C}_{11}+\tilde C_{33})V_{3;1}^{(1)}\nonumber \\
&&-2\Omega_0\omega_\theta V_{2;3}^{(1)}-2\Lambda\omega_\theta \alpha_2^{-1} V_{3;2}^{(1)}=-M\tilde{C}_{13}^{(1)}\frac{a_3^2+a_1^2}{5},~~\\
        &&(-\omega_\theta^2+4\pi\rho B_{23}-\Omega_0^2+\tilde{C}_{22}+\tilde C_{33})V_{2;3}^{(1)}\nonumber \\
        &&+(-\omega_\theta^2+4\pi\rho B_{23}-\Omega_0^2-2\Lambda^2\nonumber \\
        &&+2\Omega_0\Lambda \alpha_2^{-1}+\tilde{C}_{22}+\tilde C_{33})V_{3;2}^{(1)}\nonumber \\
        &&-2\Omega_0\omega_\theta V_{1;3}^{(1)}-2\Lambda\omega_\theta\alpha_2 V_{3;1}^{(1)}=0,
\end{eqnarray} 
while the antisymmetrized equations are 
\begin{eqnarray}
        &&(-\omega_\theta^2-\Omega_0^2+\tilde C_{11}-\tilde{C}_{33})V_{1;3}^{(1)}\nonumber \\
        &&+(\omega_\theta^2-\Omega_0^2+2\Omega_0\Lambda\alpha_2+\tilde C_{11}-\tilde{C}_{33})V_{3;1}^{(1)}\nonumber \\
        &&-2\Omega_0\omega_\theta V_{2;3}^{(1)}+2\Lambda\omega_\theta\alpha_2^{-1}V_{3;2}^{(1)}=-M\tilde{C}_{13}^{(1)}\frac{a_3^2-a_1^2}{5},\\
        &&(-\omega_\theta^2-\Omega_0^2+\tilde C_{22}-\tilde{C}_{33})V_{2;3}^{(1)}\nonumber \\
        &&+(\omega_\theta^2-\Omega_0^2+2\Omega_0\Lambda\alpha_2^{-1}+\tilde C_{22}-\tilde{C}_{33})V_{3;2}^{(1)}\nonumber \\
        &&-2\Omega_0\omega_\theta V_{1;3}^{(1)}+2\Lambda\omega_\theta\alpha_2V_{3;1}^{(1)}=0.
\end{eqnarray}

These equations can be recasted as a matrix equation 
\begin{equation}
   \sum_{k=1}^4 M_{ik}^{(1)}V_k^{(1)}=C_i^{(1)},
\end{equation} where \begin{equation}
    V_i^{(1)}=(V_{1;3}^{(1)}, V_{3;1}^{(1)}, V_{2;3}^{(1)}, V_{3;2}^{(1)}),
\end{equation}
\begin{equation}
    C_i^{(1)}=-\frac{M\tilde C_{13}^{(1)}}{5}(a_3^2+a_1^2, a_3^2-a_1^2, 0, 0),
\end{equation} 
\begin{eqnarray}
    \tilde C_{13}&=&\tilde C_{13}^{(1)}\cos(\omega_\theta\tau)\nonumber \\
    &=&3a\varepsilon_0\cos(\omega_\theta\tau)
    \frac{m\sqrt{r_0^2+K_0}(r_0^2+5K_0)}{\sqrt{K_0}r^6},
\end{eqnarray}
and 
\begin{equation}
   \begin{split}
    M^{(1)}_{11}=&-\omega_\theta^2+4\pi\rho B_{13}-\Omega_0^2+\tilde{C}_{11}+\tilde C_{33}, \\
    M^{(1)}_{12}=&-\omega_\theta^2+4\pi\rho B_{13}-\Omega_0^2-2\Lambda^2\\&+2\Omega_0\Lambda\alpha_2+\tilde{C}_{11}+\tilde C_{33}, \\
    M^{(1)}_{13}=&-2\Omega_0\omega_\theta, \\
    M^{(1)}_{14}=&-2\Lambda\omega_\theta\alpha_2^{-1}, \\
\end{split}
\end{equation}
\begin{equation}
   \begin{split}
    M^{(1)}_{21}=&-2\Omega_0\omega_\theta, \\
    M^{(1)}_{22}=&-2\Lambda\omega_\theta\alpha_2, \\
    M^{(1)}_{23}=&-\omega_\theta^2+4\pi\rho B_{23}-\Omega_0^2+\tilde{C}_{22}+\tilde C_{33}, \\
    M^{(1)}_{24}=&-\omega_\theta^2+4\pi\rho B_{23}-\Omega_0^2-2\Lambda^2\\&+2\Omega_0\Lambda\alpha_2^{-1}+\tilde{C}_{22}+\tilde C_{33}, \\
\end{split}
\end{equation}
\begin{equation}
   \begin{split}
    M^{(1)}_{31}=&-\omega_\theta^2-\Omega_0^2+\tilde C_{11}-\tilde C_{33},\\
    M^{(1)}_{32}=&\omega_\theta^2-\Omega_0^2+2\Omega_0\Lambda\alpha_2+\tilde C_{11}-\tilde C_{33},\\
    M^{(1)}_{33}=&-2\Omega_0\omega_\theta, \\
    M^{(1)}_{34}=&2\Lambda\omega_\theta\alpha_2^{-1}, \\
\end{split}
\end{equation}
\begin{equation}
   \begin{split}
    M^{(1)}_{41}=&-2\Omega_0\omega_\theta, \\
    M^{(1)}_{42}=&2\Lambda\omega_\theta\alpha_2, \\
    M^{(1)}_{43}=&-\omega_\theta^2-\Omega_0^2+\tilde C_{22}-\tilde{C}_{33}, \\
    M^{(1)}_{44}=&\omega_\theta^2-\Omega_0^2+2\Omega_0\Lambda\alpha_2^{-1}+\tilde C_{22}-\tilde{C}_{33}.\\
\end{split}
\end{equation}
Note that all the components of $M_{ij}^{(1)}$ and $C_i^{(1)}$ are written in the zeroth-order quantities, and hence, the solution for $V_k^{(1)}$ is obtained straightforwardly by inverting the matrix equation (see Sec.~\ref{Res}). 

\subsection{Second-order calculations}

To find the second-order equations, we need to take into account second-order corrections to $\Omega$ and $\dot\Omega$ for the equations. We denote the former as $\Omega=\Omega_0+\Omega^{(2)}+O(\varepsilon_0^4)$. The second-order corrections to $\Omega$ and $\dot\Omega$ are given in the appendix \ref{appendix2}.

One must not forget quadratic contributions from the first order perturbations. Indeed the total moment of inertia tensor computed up to the second order is \begin{eqnarray}
    I^{tot}_{ij}&=&\int_V(x_i+\xi^{(1)}_{ik}x_{k}+\xi^{(2)}_{ik}x_{k})(x_j+\xi^{(1)}_{jl}x_{l}+\xi^{(2)}_{jl}x_{l})d^3x\nonumber\\
    &=&I^{(0)}_{ij}+V^{(1)}_{ij}+V^{(2)}_{ij}+\frac{5}{Ma_k^2}V_{i;k}^{(1)}V^{(1)}_{j;k}+O(\varepsilon^3).
\end{eqnarray}
We will in the following abbreviate these contributions to $Q_{ij}(V^{(1)}_{k;l})$, while giving them their complete form in the appendix \ref{appendix2}. Given our displacement, these terms are non-zero only when $(i,j)$ is even on the component 3.

For the second-order perturbation, we only need to consider the even terms on the component 3. We then have five equations. By adding Eq.~(\ref{VolConserve}), we obtain six equations for six variables, $V_{1;1}$, $V_{2;2}$, $V_{3;3}$, $V_{1;2}$, $V_{2;1}$, and $\delta\Pi$.

Eliminating $\delta\Pi$ from the equations, and using the relations
\begin{multline}
    -2B_{11}V_{11}+a_1^2\sum_{l=1}^3 A_{1l}V_{ll}+2B_{33}V_{33}-a_3^2\sum_{l=1}^3 A_{3l}V_{ll}\\=-(3B_{11}-B_{13})V_{11}+(B_{23}-B_{12})V_{22}+(3B_{33}-B_{13})V_{33},
\end{multline}
 we obtain
\begin{eqnarray}
        &&\Big(\frac{d^2}{d\tau^2}+2\pi\rho (3B_{11}-B_{13})-2\Omega_0^2-2\Lambda^2+2\Omega_0\Lambda\alpha_2\nonumber \\
        &&~+2\tilde{C}_{11}\Big)V_{1;1}+(-2\pi\rho (B_{23}-B_{12})+2\Omega_0\Lambda \alpha_2^{-1})V_{2;2}\nonumber \\
        &&+\Big(-\frac{d^2}{d\tau^2}-2\pi\rho(3B_{33}-B_{13})-2\tilde C_{33}\Big)V_{3;3}\nonumber \\
        && -2\Lambda\alpha_2^{-1}\frac{d}{d\tau}V_{1;2} -2\Omega_0\frac{d}{d\tau}V_{2;1}\nonumber \\
        &&=-\tilde{C}_{11}^{(2)}I_{11}+\tilde{C}_{33}^{(2)}I_{33}+4\Omega^{(2)}(\Omega_0-\Lambda\alpha_2)I_{11}+Q_{11}(V^{(1)}_{k;l}), \nonumber \\
\end{eqnarray}
\begin{eqnarray}
        &&(-2\pi\rho (B_{13}-B_{12})+2\Omega_0\Lambda \alpha_2)V_{1;1}\nonumber\\
        &&+\Big(\frac{d^2}{d\tau^2}+2\pi\rho(3B_{22}-B_{23})-2\Omega_0^2-2\Lambda^2+2\Omega_0\Lambda\alpha_2^{-1}\nonumber \\
        &&+2\tilde{C}_{22}\Big)V_{2;2}+\Big(-\frac{d^2}{d\tau^2}-2\pi\rho(3B_{33}-B_{23})-2\tilde C_{33}\Big)V_{3;3}\nonumber \\
        &&+2\Omega_0\frac{d}{d\tau}V_{1;2}+2\Lambda \alpha_2\frac{d}{d\tau}V_{2;1}\nonumber \\
        &&=-\tilde{C}_{22}^{(2)}I_{22}+\tilde{C}_{33}^{(2)}I_{33}+\tilde{C}_{13}(V_{3;1}+V_{1;3})\nonumber \\
        &&~~+4\Omega^{(2)}(\Omega_0-\Lambda\alpha_2^{-1})I_{22}+Q_{22}(V^{(1)}_{k;l}),
\end{eqnarray}
\begin{eqnarray}
        &&(2\Lambda\alpha_2+2\Omega_0)\frac{d}{d\tau}V_{1;1}-(2\Omega_0+2\Lambda \alpha_2^{-1})\frac{d}{d\tau}V_{2;2}\nonumber\\
        &&+\Big(\frac{d^2}{d\tau^2}+4\pi\rho B_{12}-2\Omega_0^2-2\Lambda^2+\tilde{C}_{11}+\tilde C_{22}\Big)V_{1;2}\nonumber\\
        &&+\Big(\frac{d^2}{d\tau^2}+4\pi\rho B_{12}-2\Omega_0^2-2\Lambda^2+\tilde{C}_{11}+\tilde C_{22}\Big)V_{2;1}\nonumber\\
        &&=-2\tilde{C}_{13}(V_{3;2}+V_{2;3})+\dot{\Omega}(I_{22}-I_{11})\nonumber\\
        &&+Q_{12}(V^{(1)}_{k;l})+Q_{21}(V^{(1)}_{k;l}),
\end{eqnarray}
\begin{eqnarray}
        &&(2\Lambda\alpha_2-2\Omega_0)\frac{d}{d\tau}V_{1;1}-(2\Omega_0-2\Lambda \alpha_2^{-1})\frac{d}{d\tau}V_{2;2}\nonumber\\
        &&+\Big(\frac{d^2}{d\tau^2}+\tilde C_{11}-\tilde{C}_{22}\Big)V_{1;2}\nonumber\\
        &&+\Big(-\frac{d^2}{d\tau^2}+\tilde C_{11}-\tilde{C}_{22}\Big)V_{2;1}\nonumber\\
        &&=\dot{\Omega}(I_{22}+I_{11})+Q_{12}(V^{(1)}_{k;l})-Q_{21}(V^{(1)}_{k;l}),
\end{eqnarray} 
and
\begin{equation}
   V_{1;1}+\frac{1}{\alpha_2^2}V_{2;2}+\frac{1}{\alpha_3^2}V_{3;3}=\sum_{k=1}^3\sum_{l=1}^3\frac{V^{(1)}_{k;l}V^{(1)}_{l;k}}{2\alpha_k^2\alpha_l^2}.
\end{equation}

There is an additional subtlety in this second-order equations since the right hand side of the equations has two different time dependence because of  $\varepsilon^2(t)=\varepsilon_0^2[1+\cos(2\omega_\theta\tau)]/2$. That is, there is a contribution from the constant terms and from the terms of angular velocity $2\omega_\theta$.

We must treat the different time dependence separately. We thus separate the terms of different frequencies and make the ansatz
\begin{equation}
    \begin{split}
        V_{1;1}&=V_{1;1}^{(2,0)}+V_{1;1}^{(2,2)}\cos(2\omega_\theta\tau),\\ V_{2;2}&=V_{2;2}^{(2,0)}+V_{2;2}^{(2,2)}\cos(2\omega_\theta\tau),\\ V_{3;3}&=V_{3;3}^{(2,0)}+V_{3;3}^{(2,2)}\cos(2\omega_\theta\tau),\\ V_{1;2}&=V_{1;2}^{(2,0)}+V_{1;2}^{(2,2)}\sin(2\omega_\theta\tau),\\
        V_{2;1}&=V_{2;1}^{(2,0)}+V_{2;1}^{(2,2)}\sin(2\omega_\theta\tau). 
    \end{split}
\end{equation}

We then separate the source terms in the same way: 
\begin{equation}
\label{splitTimeDep}
\begin{split}
    \tilde C_{11}^{(2)}&=\tilde C_{11}^{(2,0)}+\tilde C_{11}^{(2,2)}\cos(2\omega_\theta\tau),\\
    \tilde C_{22}^{(2)}&=\tilde C_{22}^{(2,0)}+\tilde C_{22}^{(2,2)}\cos(2\omega_\theta\tau),\\
    \tilde C_{33}^{(2)}&=\tilde C_{33}^{(2,0)}+\tilde C_{33}^{(2,2)}\cos(2\omega_\theta\tau),\\
    \Omega^{(2)}&=\Omega^{(2,0)}+\Omega^{(2,2)}\cos(2\omega_\theta\tau),\\
    \dot\Omega&=\dot\Omega^{(2)}\sin(2\omega_\theta\tau),\\
    \tilde{C}_{13}(V_{3;1}+V_{1;3})&=\frac{\tilde{C}_{13}^{(1)}(V_{3;1}^{(1)}+V_{1;3}^{(1)})}{2}[1+\cos(2\omega_\theta\tau)],\\
    \tilde{C}_{13}(V_{3;2}+V_{2;3})&=\frac{\tilde{C}_{13}^{(1)}(V_{3;2}^{(1)}+V_{2;3}^{(1)})}{2}\sin(2\omega_\theta\tau).
\end{split}
\end{equation}
For the complete expression of these terms, we refer the reader to the appendix.

As before, we write the corresponding equations using the matrix notation
\begin{equation}
\sum_{k=1}^3 M_{ik}^{(2,2)}V_k^{(2,2)}=C_i^{(2,2)},
\end{equation}
where
\begin{equation}
   \begin{split}
    M^{(2,2)}_{11}=&-4\omega_\theta^2+2\pi\rho (3B_{11}-B_{13})\\&-2\Omega_0^2-2\Lambda^2+2\Omega_0\Lambda\alpha_2+2\tilde{C}_{11}, \\
    M^{(2,2)}_{12}=&-2\pi\rho (B_{23}-B_{12})+2\Omega_0\Lambda\alpha_2^{-1}, \\
    M^{(2,2)}_{13}=&4\omega_\theta^2-2\pi\rho(3B_{33}-B_{13})-2\tilde C_{33}, \\
    M^{(2,2)}_{14}=&-4\Lambda\omega_\theta\alpha_2^{-1}, \\
    M^{(2,2)}_{15}=&-4\Omega_0\omega_\theta, \\
\end{split}
\end{equation}
\begin{equation}
   \begin{split}
    M^{(2,2)}_{21}=&-2\pi\rho (B_{13}-B_{12})+2\Omega_0\Lambda\alpha_2,\\
    M^{(2,2)}_{22}=&-4\omega_\theta^2+2\pi\rho(3B_{22}-B_{23})\\&-2\Omega_0^2-2\Lambda^2+2\Omega_0\Lambda\alpha_2^{-1}+2\tilde{C}_{22},\\
    M^{(2,2)}_{23}=&4\omega_\theta^2-2\pi\rho(3B_{33}-B_{23})-2\tilde C_{33}, \\
    M^{(2,2)}_{24}=&4\Omega_0\omega_\theta, \\
    M^{(2,2)}_{25}=&4\Lambda\omega_\theta\alpha_2, \\
\end{split}
\end{equation}
\begin{equation}
   \begin{split}
    M^{(2,2)}_{31}=&(-4\Lambda\alpha_2-4\Omega_0)\omega_\theta, \\
    M^{(2,2)}_{32}=&(4\Lambda\alpha_2^{-1}+4\Omega_0)\omega_\theta, \\
    M^{(2,2)}_{33}=&0,\\
    M^{(2,2)}_{34}=&-4\omega_\theta^2+4\pi\rho B_{12}-2\Omega_0^2-2\Lambda^2+\tilde{C}_{11}+\tilde C_{22}, \\
    M^{(2,2)}_{35}=&-4\omega_\theta^2+4\pi\rho B_{12}-2\Omega_0^2-2\Lambda^2+\tilde{C}_{11}+\tilde C_{22}, \\
\end{split}
\end{equation}
\begin{equation}
   \begin{split}
    M^{(2,2)}_{41}=&(-4\Lambda\alpha_2+4\Omega_0)\omega_\theta, \\
    M^{(2,2)}_{42}=&(-4\Lambda\alpha_2^{-1}+4\Omega_0)\omega_\theta, \\
    M^{(2,2)}_{43}=&0,\\
    M^{(2,2)}_{44}=&-4\omega_\theta^2+\tilde C_{11}-\tilde{C}_{22}, \\
    M^{(2,2)}_{45}=&4\omega_\theta^2+\tilde C_{11}-\tilde{C}_{22},\\
\end{split}
\end{equation}\begin{equation}
   \begin{split}
    M^{(2,2)}_{51}=&1, \\
    M^{(2,2)}_{52}=&\frac{1}{\alpha_2^2}, \\
    M^{(2,2)}_{53}=&\frac{1}{\alpha_3^2}, \\
    M^{(2,2)}_{54}=&0, \\
    M^{(2,2)}_{55}=&0.\\
\end{split}
\end{equation} $V^{(2,2)}=(V_{1;1}^{(2,2)}, V_{2;2}^{(2,2)}, V_{3;3}^{(2,2)}, V_{1;2}^{(2,2)}, V_{2;1}^{(2,2)})$, and \begin{equation}
    \begin{split}
        C^{(2,2)}_1=&-\tilde{C}_{11}^{(2,2)}I_{11}+\tilde{C}_{33}^{(2,2)}I_{33}\\&+4\Omega^{(2,2)}(\Omega_0-\Lambda\alpha_2)I_{11}+Q_{11}^{(2,2)}(V^{(1)}_{k;l}), \\
        C^{(2,2)}_2=&-\tilde{C}_{22}^{(2,2)}I_{22}+\tilde{C}_{33}^{(2,2)}I_{33}+\frac{\tilde{C}_{13}^{(1)}(V_{3;1}^{(1)}+V_{1;3}^{(1)})}{2}\\&+4\Omega^{(2,2)}(\Omega_0-\Lambda\alpha_2^{-1})I_{22}+Q_{22}^{(2,2)}(V^{(1)}_{k;l}), \\
        C^{(2,2)}_3=&-\tilde{C}_{13}^{(1)}(V_{3;2}^{(1)}+V_{2;3}^{(1)})+\dot{\Omega}^{(2)}(I_{22}-I_{11})\\&+Q_{12}^{(2,2)}(V^{(1)}_{k;l})+Q_{21}^{(2,2)}(V^{(1)}_{k;l}), \\
        C^{(2,2)}_4=&\dot{\Omega}(I_{22}+I_{11})+Q_{12}^{(2,2)}(V^{(1)}_{k;l})-Q_{21}^{(2,2)}(V^{(1)}_{k;l}), \\
        C^{(2,2)}_5=&\sum_{k=1}^3\sum_{l=1}^3\frac{V^{(1)}_{k;l}V^{(1)}_{l;k}}{2\alpha_k^2\alpha_l^2}. \\
    \end{split}
\end{equation}

The differential equations for the constant term must be considered with more care. If we proceed as before, we obtain again linear equations $\displaystyle \sum_{k=1}^5 M_{ik}^{(2,0)}V_k^{(2,0)}=C_i^{(2,0)}$. The matrix of this equation $M_{ik}^{(2,0)}$ is the same as $M_{ik}^{(2,2)}$ when we set $\omega_\theta=0$. However, this matrix is never invertible as its third and fourth line are colinear. 

However, in this case the term in the right hand side, $C_i^{(2,0)}$, is also simpler because $C_3^{(2,0)}=C_4^{(2,0)}=0$ (see Eq.~(\ref{splitTimeDep})). This would imply that we have four equations for five variables. We have thus a certain freedom for setting $V_{1;2}^{(2,0)}$ and $V_{2;1}^{(2,0)}$. We could set them both to 0.

By exploring further, we also find that one can assume $V_{1;2}^{(2,0)}=w_1\tau$, $V_{2;1}^{(2,0)}=w_2\tau$. The differential equations only impose $w_1+w_2=0$.
It thus seems that there is a certain freedom on the choice of the angular momentum at this order. While we could fix all these variables to 0 as a simplifying assumption, we could also enforce the value $f$, as defined in Eq.~(\ref{f}), at this order by comparing the corresponding vorticity of this second-order correction to the second-order correction of $\Omega$. We may then set 
\begin{equation}
    w_1=\frac{2a_1^2a_2^2\Omega^{(2,0)}f}{a_1^2+a_2^2}.
\end{equation}

\subsection{Higher orders}

Even for higher orders, the form of the equations stays essentially the same (the matrix form is identical). Indeed, as before, we have linear equations for higher-order displacements. 

For higher orders, however, there are three main differences from the previous analyses. The first comes from the right hand side of the matrix equations. It then reflects the contribution of all the lower-order displacements and the higher-order contributions of the physical parameters of the system. The second comes from the different frequencies which one has to consider: Specifically, the frequency is higher. Finally, higher-order contributions to the geodesic motion have to be taken into account.

\subsection{Elliptic orbits}

The method described in the previous subsections can be extended to other types of orbits, like slightly elliptic orbits. For this, we write the orbital separation as $r(\tau)=r_0[1+\varepsilon(\tau)]$ and assume that $\varepsilon(\tau)$ is an infinitesimal quantity.

As before, we need to determine the time-dependence of $\varepsilon$. 
Using the definition of $\varepsilon$ and writing $E$, $L$ and $K$ by the geodesic constants of a circular geodesic of radius $r_0$, we get from Eq.~(\ref{Req}) as 
\begin{equation}
    \left(\frac{d\varepsilon}{d\tau}\right)^2=(\delta r)^2-\varepsilon^2\Omega_0^2\left[1-\frac{3(L-aE)^2}{r_0^2}\right],
\end{equation} 
where $\delta r$ is given by the perturbed values of the geodesic constants, $E$ and $L$, and $K=(L-aE)^2$ for the equatorial orbits. We then obtain
\begin{equation}
    \varepsilon(\tau)=\delta r\cos(\omega_e\tau), 
\end{equation} 
where 
\begin{eqnarray}
\omega_e=\Omega_0\sqrt{1-\frac{3(L-aE)^2}{r_0^2}} 
=\Omega_0 \sqrt{4-3{\Delta_0 \over P_0}}.
\end{eqnarray}
In this setting, the tidal tensor throughout the orbit is rewritten in the perturbative form. 


We can also consider more general orbits, which are neither circular nor equatorial. In this case, when we are at order $n$, we need to consider all frequencies $l\omega_\theta+m\omega_e$ for all $|l|+|m|=n$, with $l,m\in\mathbf{Z}$. This is a topic beyond the scope of our present work. 

\begin{figure}[t]
     \centering
         \includegraphics[width=0.5\textwidth]{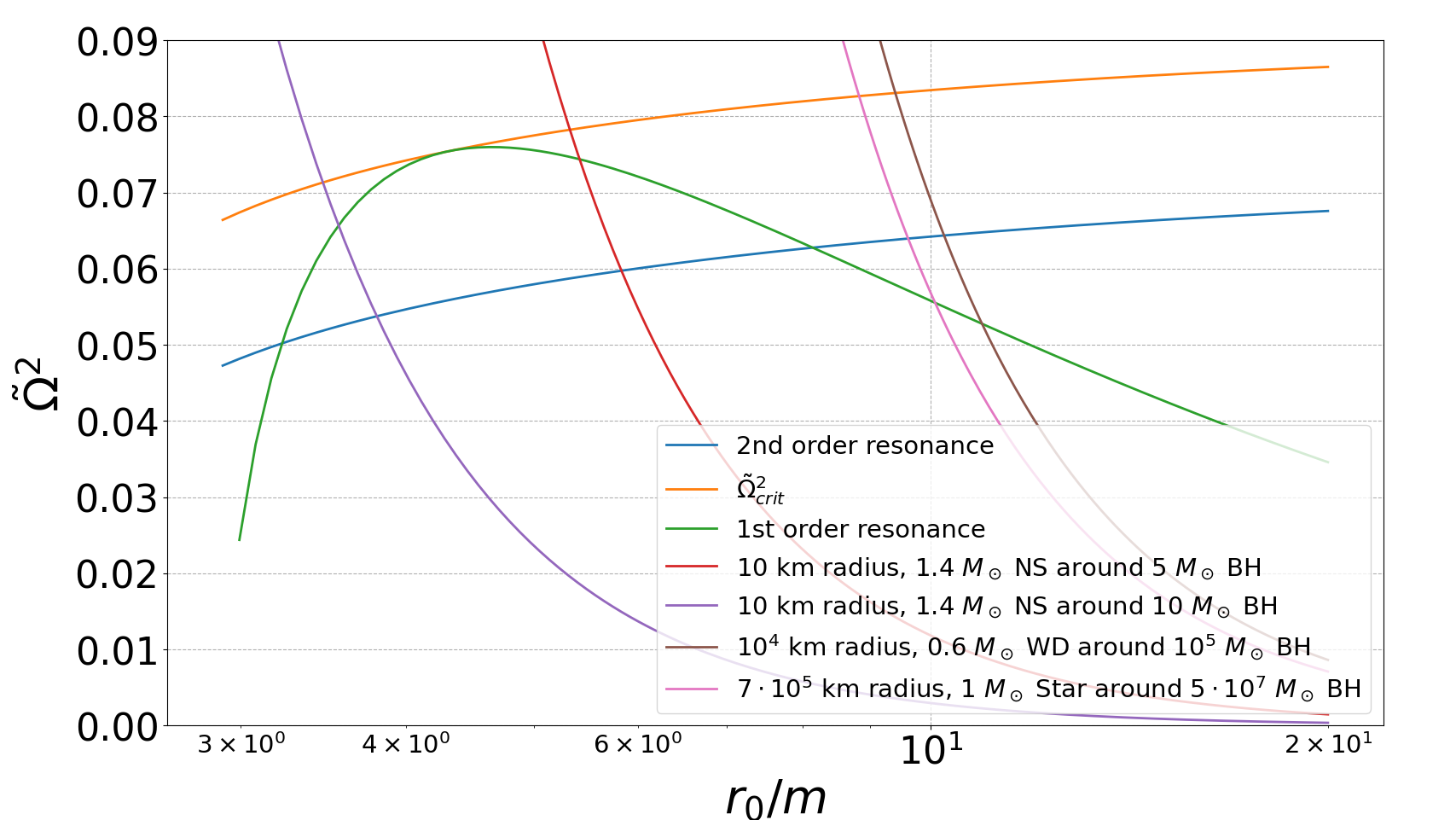}
     \hfill
         \includegraphics[width=0.5\textwidth]{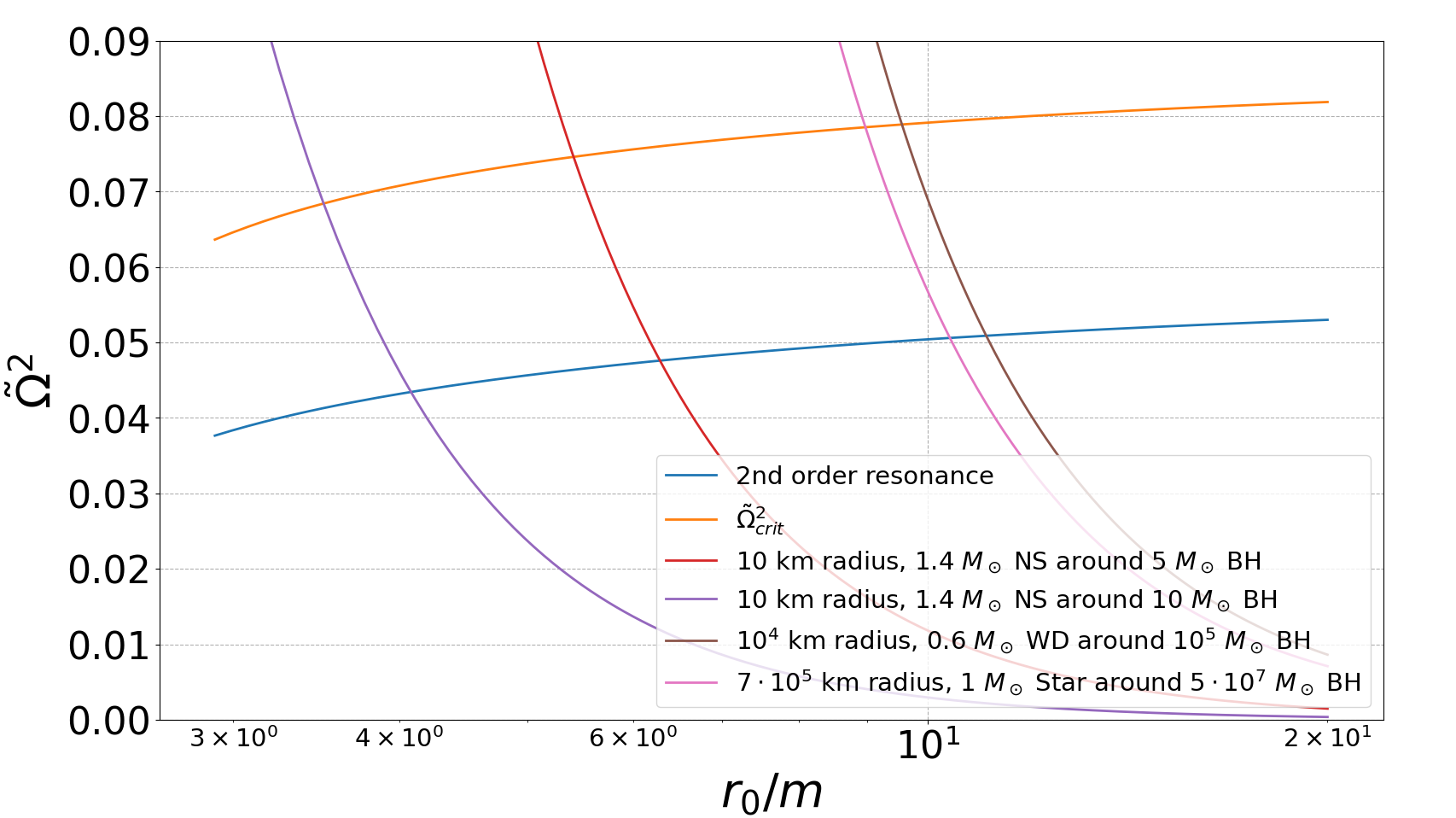}
     \caption{\label{DistRes} $\tilde\Omega^2$ as functions of $r_0/m$ for the first- and second-order resonances, along with the tidal disruption limit $\tilde\Omega^2_\mathrm{crit}$ as well as the expected values of $\tilde\Omega^2$ for typical values of BH-NS binaries and supermassive BH-white dwarf/ordinary star binaries. For both plots, we assumed $a/m=0.8$ and prograde orbits. The upper and lower panels show the results for $f=0$ and $f=1$.}
\end{figure}

\begin{figure}[t]
     \centering
         \includegraphics[width=0.5\textwidth]{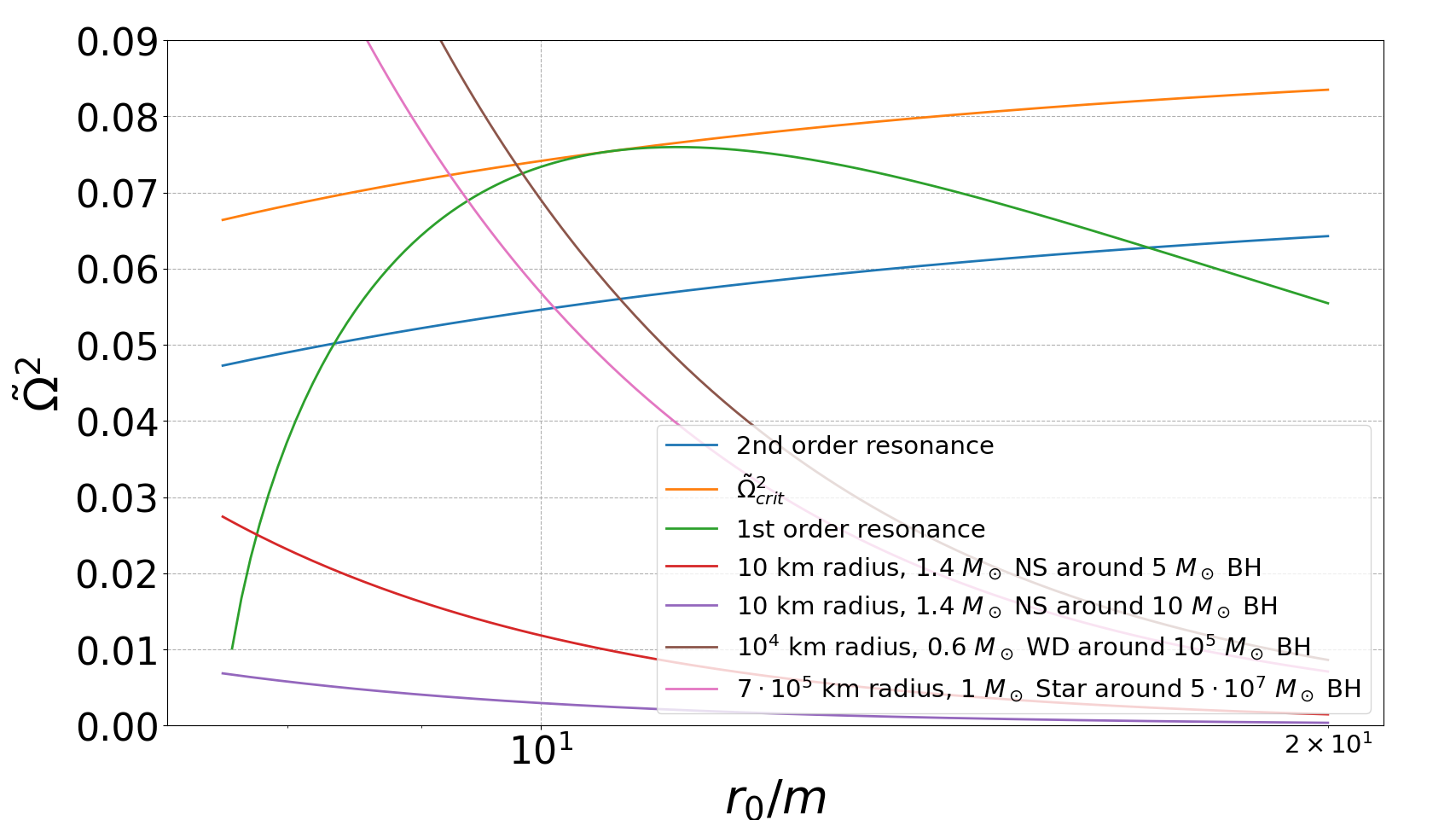}
     \hfill
         \includegraphics[width=0.5\textwidth]{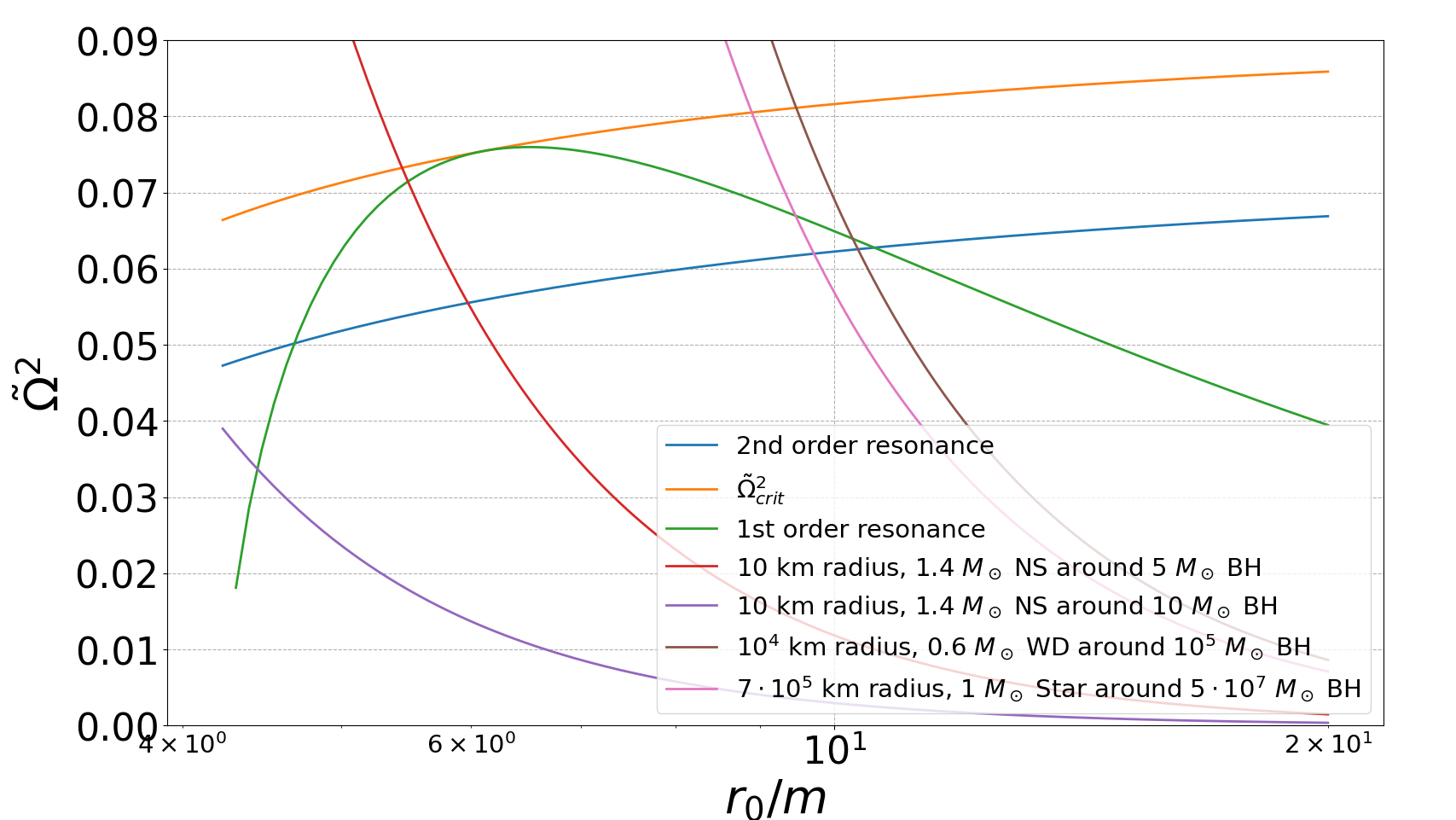}
     \hfill
         \includegraphics[width=0.5\textwidth]{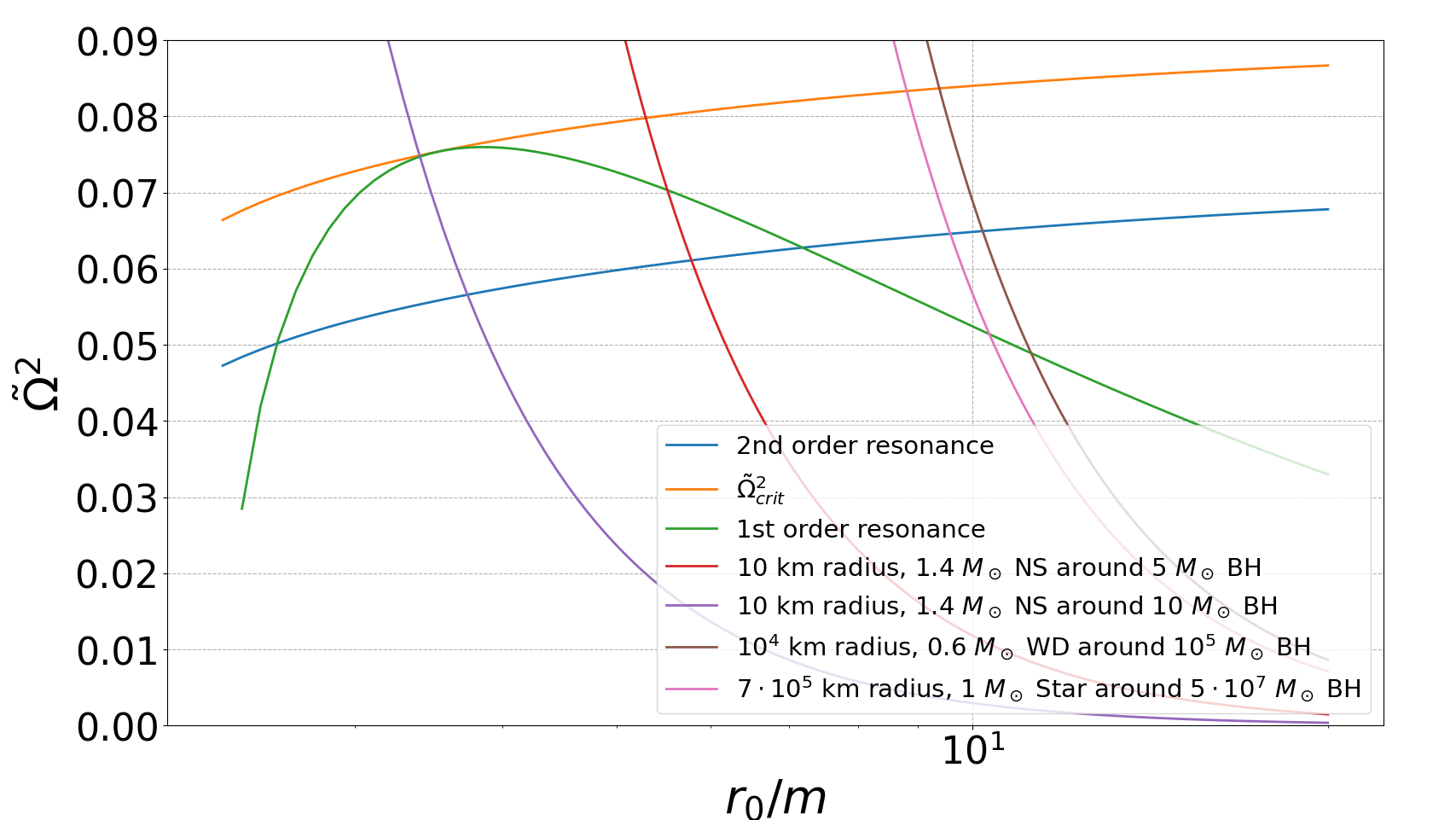}
     \caption{\label{Different_a} 
     The same as Fig.~\ref{DistRes} but for $a/m=0.5$ (upper and middle panels) and for $a/m=0.9$ (lower panel). The upper panel shows the case of retrograde orbits, while the middle and lower panels prograde orbits.}
\end{figure}

\section{Resonances}\label{Res}

\subsection{Determination of resonances}

An important aspect of the linear equations described above comes from the ansatz we made about the time dependence of the displacement. These are justified as long as the matrix equations are solvable. This hinges on the inversability of the matrices considered.

However, during our research, we found that these matrices could have determinant zero for particular equilibrium states. Furthermore, the terms in the right hand side were not in the image of the matrices. Thus the linear equations were not solvable for the particular cases in which the determinant vanishes, implyimg the presence of a resonance. This indicate that the ansatz used in these cases are inappropriate. 
To solve the equations at the resonance point, a better ansatz would be: $\vec\xi=(\vec\xi_0+\vec\xi_1\tau)e^{i\omega_\theta\tau}$. In this case, the displacements would eventually not be infinitesimal. Thus our perturbative expansion breaks down at such points. 

\begin{figure}[t]
     \centering
         \includegraphics[width=0.35\textwidth]{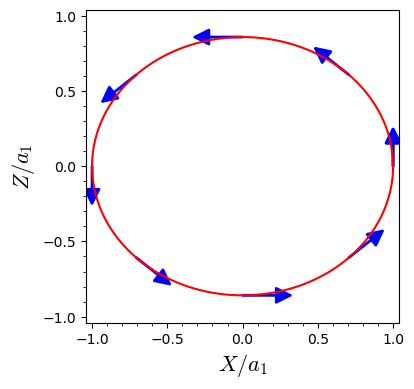}
         \includegraphics[width=0.35\textwidth]{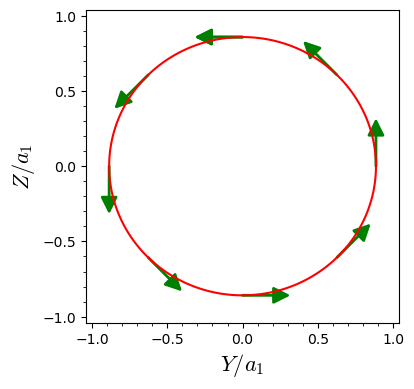}
     \caption{\label{1st} The displacement vector $\xi_i$ for the first-order perturbation on the surface of the star represented by the red ellipse for $\omega_\theta\tau=0$ (upper panel) and $\omega_\theta\tau=\pi/2$ (lower panel).
     The axes $X$, $Y$, and $Z$ correspond to the axes 1, 2, and 3 of the problem in units of $a_1$, respectively. For all the plots, the vectors are plotted for $a/m=0.8$, $f=0$ and at a distance $r_0=20\,m$ with $\tilde \Omega^2\approx 0.03459$. In this case, $\alpha_2\approx0.886$ and $\alpha_3\approx0.858$.}
\end{figure}

\begin{figure}[t]
     \centering
         \includegraphics[width=0.36\textwidth]{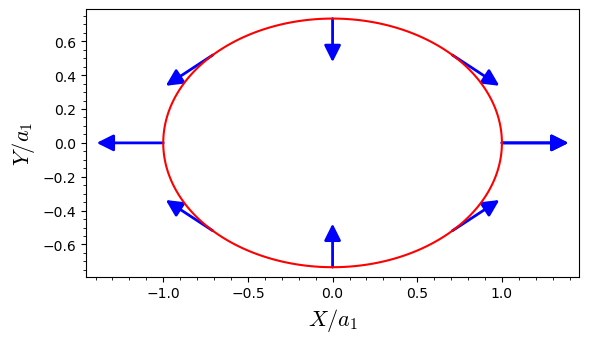}
\includegraphics[width=0.36\textwidth]{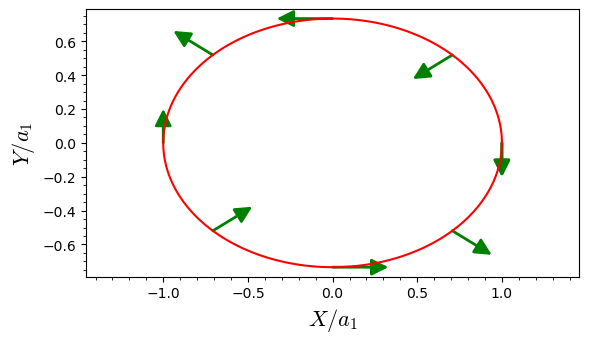}
    \caption{\label{2nd} The displacement vector $\xi_i$ for a second-order perturbation on the surface of the star represented by the red ellipse for $\omega_\theta\tau=0$ (upper panel) and $\omega_\theta\tau=\pi/2$ (lower panel).
     The axes $X$, $Y$, and $Z$ correspond to the axes 1, 2, and 3 of the problem in units of $a_1$, respectively. For all the plots, the vectors are plotted for $a/m=0.8$, $f=0$ and at a distance $r_0=20\,m$ with $\tilde \Omega^2\approx 0.06758$. In this case, $\alpha_2\approx0.734$ and $\alpha_3\approx0.694$.}
\end{figure}

These resonances can come either from the first- or second-order equation. For $f=0$, we find a resonance for both the first and second-order displacements. For $f=1$, we find it only for the second-order equations. Near the resonant angular frequencies, $\omega\approx \omega_\theta$ or $2\omega_\theta$, the amplitude of the displacement scales as $|\omega-k\omega_\theta|^{-1}$ where $k=1$ or $2$,  indicating that a significant deformation from the ellipsoidal shape happens not only at the resonant angular frequency but also around these frequencies, i.e., at $\omega$ that satisfies $|\omega-k\omega_\theta| \ll k\omega_\theta$. This indicates that for a star approaching the resonance, the degree of the displacement is significantly enhanced.

In order to assess the relevancy of these resonances, we plot the corresponding values of $\tilde\Omega^2$ of these resonances, along with the values of $\tilde\Omega^2$ that are modeled by an NS, a white dwarf, and an ordinary star orbiting a variety of Kerr BHs. 
We estimate $\tilde\Omega^2$ using typical radius $R_\mathrm{star}$ and mass $M$ for these objects. We then have $\tilde\Omega^2$ as a function of $r_0$ as
\begin{equation}
    \tilde\Omega^2=\frac{4}{3}\frac{m}{M}\left(\frac{R_\mathrm{star}}{r_0}\right)^3.
\end{equation} 

In Fig.~\ref{DistRes}, we plot the values of $\tilde \Omega^2$ for the resonances of $\omega=\omega_\theta$ and $2\omega_\theta$ as functions of $r_0/m$ together with the Roche $(f=0)$ and Roche-Riemann $(f=1)$ limits, for $a/m=0.8$. We also plot the curves along hypothetical sequences of inspiraling BH-NS binaries in close orbits and white dwarf/ordinary stars orbiting supermassive BHs. We note that for the $\omega=0$ mode in the second-order displacement, the resonance appears near the Roche/Roche-Riemann limits. This corresponds to the onset of dynamical instability for the ellipsoid, similarly as in the classical Roche problem \cite{CS}. This figure shows that these resonances may destabilize the objects before reaching the Roche and Roche-Riemann limits. 

In Fig.~\ref{Different_a}, we reproduce the same figure for $a/m=0.5$ and $a/m=0.9$ with $f=0$. We find that the spin parameter $a/m$ does not have much of an impact on the relative position of the resonances to the Roche limit. As in \cite{Sh}, we see that the Roche limit gets further away and with it the resonances, as $a/m$ decreases. 

In the upper panel of Fig.~\ref{Different_a}, we produced the curves in the case of retrograde orbits. While until now we always presented the results for prograde orbits, they are easily extended to retrograde orbits. One only has to take $a/m<0$. The observation made before can then also be extended. Compared to positive values of $a/m$, we observe that the resonances happen for larger values of $r_0/m$. However their relative positions to the Roche limit are the same.


These resonances seem to be similar to the instabilities found for a star in elliptic orbits~\cite{NewtEllipt} (note that these are instabilities in purely Newtonian gravity). They appear when $\omega_e$ is equal to one of the oscillatory modes of the star on the equatorial plane. This then creates the resonance discussed before. In the case of a star orbiting a Kerr BH, the possible resonances are more numerous as a result of the more complicated nature of the geodesics due to the relativistic effects.

In our present context, the displacement associated with an oscillatory mode gets excited by the variation of the $\theta$-coordinate of the orbit. It will then grow linearly and eventually, it will not be infinitesimal anymore. Figures~\ref{1st} and \ref{2nd} show the displacement vectors associated with the excited modes, assuming $a/m=0.8$. Again the spin parameter does not have an important impact on the qualitative aspect of the resonant displacement.

Fig.~\ref{1st} shows that the resonance in the first-order displacement will mostly imply a precession of the 3-axis. As such, this mode is quite reminiscent of the transverse-shear mode of the MacLaurin spheroids discussed in~\cite{CS}. By the non-linear amplification of this mode, matter may be ejected from the surface to the outside of the star, resulting possibly in the modification of the precessing orbit as well as the stellar profile. For white dwarfs, the stellar radius increases with the mass ejection. Thus, this resonance could trigger the tidal disruption.

From the second-order displacement shown in Fig.~\ref{2nd}, we observe that it will involve mostly a quadrupole displacement in the 1-2-plane. This mode reminds us of toroidal modes also discussed in~\cite{CS}. 

The resonance in the first- and second-order displacements is always encountered slightly outside the Roche/Roche-Riemann limits. For BH-NS binaries in precessing orbits, thus, the resonant deformation may be excited just prior to the tidal disruption. The tidal disruption may be assisted by the prior resonance if the relativistic effect on the self-gravity of the star, which we do not take into account in the present study, does not qualitatively change the property of the resonance. 


We note that in our analysis we neglected several effects, which can modify our results quantitatively. Since we neglected the general relativistic effect for the self-gravity of the NS, an error of order $Gm_\mathrm{NS}/c^2 R_\mathrm{star}$ may be expected on the condition of the tidal disruption. Also we neglected the NS's gravity on the orbital motion, so that an error of order $M/m$ may be also expected. However, these dimensionless parameters are supposed to be of order 0.1 in this paper. Hence, the results obtained in this paper is unlikely to be significantly changed even if we take into account these effects. 

\subsection{Analytical resolution of the first-order resonance}\label{secIVB}

The resonance in the first-order displacement with $f=0$ can be determined analytically. 
If we introduce $\lambda$ such that $\omega_{\theta}^2=(1+\lambda)\Omega_0^2$ where $\lambda$ is a purely relativistic correction (see Eq.~(\ref{omegat})), we have $\tilde C_{11}=(-2-\lambda)\Omega_0^2$, $\tilde C_{22}=\Omega_0^2$, and $\tilde C_{33}=(1+\lambda)\Omega_0^2$. Then, 
\begin{eqnarray}
    \det(M_{ij}^{(1)})&=&-16\pi\rho B_{23}(1+\lambda)\Omega_0^4
    \nonumber \\
    &&\times
    \Big[-4\pi\rho B_{13}\lambda+(3+\lambda)(1+\lambda)\Omega_0^2\Big],~~~
\end{eqnarray}
and thus, we find that the first-order resonance occurs when
\begin{equation}
    \tilde\Omega^2=\frac{4B_{13}\lambda}{(3+\lambda)(1+\lambda)}.
\end{equation}
We find that the relativistic correction is the key for this resonance. 

\section{Discussion}\label{Dis}

The main result of this paper is the discovery of possible resonances of stars with precessing orbits around Kerr BHs. In particular, we found that these resonances could be relevant for systems such as white dwarfs/ordinary stars orbiting a supermassive BH and BH-NS binaries in close orbits. 

One interpretation of our results is that as an orbiting star approaches a Kerr BH, e.g., by the radiation reaction of gravitational waves, it will approach one of these resonances before encountering the tidal disruption limit. The closer it gets to it, the more the displacement associated with it will grow. At some point before the resonance, the displacement will not be infinitesimal anymore. Our equations are then not valid.

As such, the instability found is unlike that found at the Roche limit. Indeed, we do not provide a correction to this limit as in \cite{TidalNonEq}. Instead, we provide a resonance which arises due to the dynamical nature of our analysis. Thus our results and their interpretation differ from \cite{TidalNonEq}.

As the first-order resonance has only been found for $f=0$, it will probably not be relevant for BH-NS binaries for which the NS is likely to be nearly irrotational~\cite{noVisc_BHNS, noVisc_BNS} in such systems. 
If we consider white dwarfs/ordinary stars orbiting supermassive BH, we expect that the corresponding star may be corotational because the viscous angular momentum redistribution can have a timescale shorter than the orbital evolution one. For these systems, the star may encounter the first-order resonance before tidal disruption.
If the system has an elliptic orbit, we expect also resonances of frequency $\omega_e$. However, if the object stays on the equatorial plane, the modification to the orbit will only give contributions to the diagonal components of the tidal tensor. As such, the resonances should correspond to the toroidal mode. 
If the orbit is neither equatorial nor spherical, the overall picture gets more complicated. According to our perturbation analysis, it is likely that there are more possible resonances of frequency $\omega_\theta+\omega_e$ and $\omega_\theta-\omega_e$, and thus, these resonances may be excited as well. 

The second-order resonance can be relevant for BH-NS binaries just prior to the significant tidal deformation of the NS. In the presence of a precession, this resonance is excited during the late inspiral phase, and thus, the orbital energy is partly transported to the oscillation energy of the NS. This can accelerate the orbital evolution of the system, leading to a gravitational-wave phase shift, in addition to the phase shift related to the usual tidal deformation~\cite{Lai:1993pa, Flanagan:2007ix}. Since the second-order resonance can be induced only at an orbit close to the tidal disruption limit, it may not be easy to distinguish the resonant effect from the usual tidal deformation effect, and thus, we may overestimate the tidal deformation effect if we do not properly take into account the resonant effect for precessing binaries.


To our knowledge, these results have never been reported before. A major counterpoint to our analysis is its apparent reliance on some questionable assumptions, like the incompressibility hypothesis. However, as the general mechanism for the resonance holds true, we expect resonances even with more realistic equations of state. Indeed, whatever the equation of state is, we have characteristic modes for our star. We may then observe a resonance between these modes and the oscillation present in the tidal tensor. Thus, while the equations can change, the resulting phenomenon should not qualitatively. Of course, this explanation is just tentative and should be confirmed by a more complete analysis of the question. We plan to show the results of our approximate analysis for the compressible equations of state in a follow-up paper (Stockinger and Shibata, in preparation).


As our equations are not valid near the resonance, one would have to change our method to fully understand the effects on the orbital change by the resonance. For this purpose, performing hydrodynamics simulations is a robust approach. 
We are currently working on simulations of these types of systems (Lam et al., in preparation).

\begin{acknowledgments}

We thank the member of Computational Relativistic Astrophysics group at the Max Planck Institute for Gravitational Physics for discussions.  
This work was in part supported by Grant-in-Aid for Scientific Research (grant Nos.~20H00158 and 23H04900) of Japanese MEXT/JSPS.

\end{acknowledgments}

\appendix

\section{Formula for second-order equations}
\label{appendix}

In order to compute all the second-order quantities necessary for the corresponding equations, we need to compute the perturbed expressions for the energy $E$, angular momentum around the $z$-axis $L$, and the Carter constant $K$ for the corresponding orbit.

To do this we will rely on the formula given in \cite{Spherical}.
Assuming that $\cal{C}$ is a small quantity, we expand them as
\begin{eqnarray}
    E&=&E_0+E_1\mathcal{C}+O(\mathcal{C}^2),\\
    L&=&L_0+L_1\mathcal{C}+O(\mathcal{C}^2),\\
    K&=&K_0+K_1\mathcal{C}+O(\mathcal{C}^2),
\end{eqnarray} 
where 
\begin{widetext}
\begin{eqnarray}
    E_1&=&\frac{a}{2r_0^2}\frac{a-\sqrt{mr}}{\sqrt{P_0}},\\
    L_1&=&\frac{a^3\sqrt{m}+a\sqrt{m}r_0^2+r_0^{5/2}(-3m+r_0)+a^2r_0^{1/2}(-m+r_0)}{2\sqrt{mP_0}r_0^5},\\
    K_1&=&1-\sqrt{\frac{r_0}{m}}\frac{(-a+\sqrt{mr_0})(a^2+a\sqrt{mr_0}+(-3m+r_0))}{P_0}.
\end{eqnarray}
\end{widetext}
Then, we have
\begin{widetext}
\begin{eqnarray}
    \Omega^{(2)}&=&
    a^2\cos^2\theta\frac{\sqrt{m}(r_0^2-3mr_0+2a\sqrt{mr_0})}{r_0^{7/2}(\sqrt{mr_0}-a)^2}\nonumber \\
    &&+\mathcal{C}\left[2K_1\Omega_0+\frac{\sqrt{K_0}}{r_0^2}\left(\frac{E_1(r_0^2+a^2)-aL_1}{r_0^2+K_0}-\frac{K_1(E_0(r_0^2+a^2)-aL_0)}{(r_0^2+K_0)^2}+\frac{a(L_1-aE_1)}{K_0}-\frac{aK_1(L_0-aE_0)}{K_0^2}\right)\right]\nonumber\\
    &=&a^2\cos^2\theta\Omega_a+\mathcal{C}\Omega_\mathcal{C}.
\end{eqnarray}
\end{widetext} 
Therefore 
\begin{equation}
    \begin{split}
        \Omega^{(2,0)}&=\frac{a^2\varepsilon_0^2}{2}\Omega_a+\mathcal{C}\Omega_\mathcal{C},\\
        \Omega^{(2,2)}&=\frac{a^2\varepsilon_0^2}{2}\Omega_a.
    \end{split}
\end{equation}

Likewise, we have  
\begin{equation}
    \dot\Omega=-a^2\dot\theta\sin2\theta\frac{\sqrt{m}(r_0^2-3mr_0+2a\sqrt{mr_0})}{r_0^{7/2}(\sqrt{mr_0}-a)^2}+O(\mathcal{C}^2),
\end{equation}
and thus
\begin{equation}
    \dot\Omega^{(2)}=-a^2\varepsilon_0^2\omega_\theta\frac{\sqrt{m}(r_0^2-3mr_0+2a\sqrt{mr_0})}{r_0^{7/2}(\sqrt{mr_0}-a)^2}.
\end{equation}

Finally, the second-order contributions to the tidal tensor are given as
\begin{equation}
    C_{11}^{(2)}=\frac{3M}{r_0^3}\left(-\frac{K_1\mathcal{C}}{r_0^2}+a^2\cos^2\theta\left(\frac{15K_0^2+14r_0^2K_0+r_0^4}{r_0^4K_0}\right)\right),
\end{equation}
\begin{equation}
    C_{22}^{(2)}=-\frac{6M}{r_0^5}a^2\cos^2\theta,
\end{equation}
\begin{equation}
    C_{33}^{(2)}=\frac{3M}{r_0^3}\left(\frac{K_1\mathcal{C}}{r_0^2}-a^2\cos^2\theta\left(\frac{15K_0^2+12r_0^2K_0+r_0^4}{r_0^4K_0}\right)\right).
\end{equation}
Thus
\begin{equation}
    \begin{split}
    C_{11}^{(2,0)}&=\frac{3M}{r_0^3}\left(-\frac{K_1\mathcal{C}}{r_0^2}+\frac{a^2\varepsilon_0^2}{2}\left(\frac{15K_0^2+14r_0^2K_0+r_0^4}{r_0^4K_0}\right)\right),\\
    C_{11}^{(2,2)}&=\frac{3Ma^2\varepsilon_0^2}{2r_0^3}\frac{15K_0^2+14r_0^2K_0+r_0^4}{r_0^4K_0},\\
    C_{22}^{(2,0)}&=-\frac{3Ma^2\varepsilon_0^2}{r_0^5},\\
    C_{22}^{(2,2)}&=-\frac{3Ma^2\varepsilon_0^2}{r_0^5},\\
    C_{33}^{(2,0)}&=\frac{3M}{r_0^3}\left(\frac{K_1\mathcal{C}}{r_0^2}-\frac{a^2\varepsilon_0^2}{2}\left(\frac{15K_0^2+12r_0^2K_0+r_0^4}{r_0^4K_0}\right)\right),\\
    C_{33}^{(2,2)}&=\frac{3Ma^2\varepsilon_0^2}{2r_0^3}\frac{15K_0^2+12r_0^2K_0+r_0^4}{r_0^4K_0}.
    \end{split}
\end{equation}

\section{Quadratic contributions to the second order}
\label{appendix2}

The quadratic contributions to the second order can be divided in three parts \begin{eqnarray}
 Q_{ij}(V_{k;l})=\frac{5}{M}(Q_{ij}^{\text{grav}}(V_{k;l})+Q_{ij}^{\text{stat}}(V_{k;l})+Q_{ij}^{\text{dyn}}(V_{k;l})),\nonumber\\   
\end{eqnarray}

where \begin{eqnarray}
    Q_{ij}^{\text{stat}}(V_{k;l})=\sum_{k=1}^3\sum_{l=1}^3(\Omega_0^2(\delta_{ik}-\delta_{3k})-\tilde C^{(0)}_{ik})\frac{V_{k;l}V_{j;l}}{a_l^2},\nonumber\\
\end{eqnarray}

\begin{eqnarray}
    Q_{ij}^{\text{dyn}}(V_{k;l})&=&\sum_{k=1}^3\bigg(-\frac{\ddot{V}_{i;k}V_{j;k}}{a_k^2}+2\sum_{l=1}^3\big(Q_{jl}\dot{V}_{i;k}\nonumber\\
    &&-(Q^2)_{jl}V_{i;k}+(Q^2)_{il}V_{j;k}\big)\frac{V_{l;k}}{a_k^2}\nonumber\\
    &&-2\sum_{l=1}^3\epsilon_{il3}\Omega_0\big(\dot{V}_{l;k}\frac{V_{j;k}}{a_k^2}\nonumber\\
    &&+\sum_{m=1}^3(Q_{lm}V_{j;k}-Q_{jm}V_{l;k})\frac{V_{m;k}}{a_k^2}\big)\bigg)\nonumber\\
\end{eqnarray}

and finally \begin{eqnarray}        
    Q_{ij}^{\text{grav}}(V_{k;l})&=&\sum_{l=1}^3\big(\frac{\mathcal{A}_iV_{i;l}V_{j;l}}{a_l^2}+\frac{\mathcal{A}_{ij}V_{i;l}V_{j;l}a_j^2}{a_l^2}+\mathcal{A}_{ilj}V_{il}V_{lj}a_j^2\nonumber\\
    &&-V_{il}\mathcal{A}_{il}V_{lj}+V_{ll}\mathcal{A}_{il}V_{ij}+\frac{V_{ll}V_{ij}\mathcal{A}_{ijl}a_j^2}{2}\nonumber\\
    &&+\delta_{ij}\sum_{k=1}^3(-\frac{\mathcal{A}_iV_{l;k}^2a_j^2}{a_l^2a_k^2}-\frac{V_{lk}^2\mathcal{A}_{lki}a_j^2}{4}\nonumber\\
    &&-\frac{\mathcal{A}_{li}V_{l;k}^2a_j^2}{2a_k^2}+\frac{\mathcal{A}_{lki}V_{ll}V_{kk}a_j^2}{8})\big).
\end{eqnarray}

When we consider a displacement $V_{i;j}$ whose non-zero terms have indices $(i,j)=(1,3),(3,1),(2,3),(3,2)$, the quadratic contributions are non-zero only for the indices even on the component 3.

\newpage
\bibliography{ref}

\begin{thebibliography}{21}%
\makeatletter
\providecommand \@ifxundefined [1]{%
 \@ifx{#1\undefined}
}%
\providecommand \@ifnum [1]{%
 \ifnum #1\expandafter \@firstoftwo
 \else \expandafter \@secondoftwo
 \fi
}%
\providecommand \@ifx [1]{%
 \ifx #1\expandafter \@firstoftwo
 \else \expandafter \@secondoftwo
 \fi
}%
\providecommand \natexlab [1]{#1}%
\providecommand \enquote  [1]{``#1''}%
\providecommand \bibnamefont  [1]{#1}%
\providecommand \bibfnamefont [1]{#1}%
\providecommand \citenamefont [1]{#1}%
\providecommand \href@noop [0]{\@secondoftwo}%
\providecommand \href [0]{\begingroup \@sanitize@url \@href}%
\providecommand \@href[1]{\@@startlink{#1}\@@href}%
\providecommand \@@href[1]{\endgroup#1\@@endlink}%
\providecommand \@sanitize@url [0]{\catcode `\\12\catcode `\$12\catcode `\&12\catcode `\#12\catcode `\^12\catcode `\_12\catcode `\%12\relax}%
\providecommand \@@startlink[1]{}%
\providecommand \@@endlink[0]{}%
\providecommand \url  [0]{\begingroup\@sanitize@url \@url }%
\providecommand \@url [1]{\endgroup\@href {#1}{\urlprefix }}%
\providecommand \urlprefix  [0]{URL }%
\providecommand \Eprint [0]{\href }%
\providecommand \doibase [0]{https://doi.org/}%
\providecommand \selectlanguage [0]{\@gobble}%
\providecommand \bibinfo  [0]{\@secondoftwo}%
\providecommand \bibfield  [0]{\@secondoftwo}%
\providecommand \translation [1]{[#1]}%
\providecommand \BibitemOpen [0]{}%
\providecommand \bibitemStop [0]{}%
\providecommand \bibitemNoStop [0]{.\EOS\space}%
\providecommand \EOS [0]{\spacefactor3000\relax}%
\providecommand \BibitemShut  [1]{\csname bibitem#1\endcsname}%
\let\auto@bib@innerbib\@empty
\bibitem [{\citenamefont {Aizenman}(1968)}]{Aizenman}%
  \BibitemOpen
  \bibfield  {author} {\bibinfo {author} {\bibfnamefont {M.~L.}\ \bibnamefont {Aizenman}},\ }\bibfield  {title} {\bibinfo {title} {The equilibrium and the stability of the roche-riemann ellipsoids},\ }\href {https://api.semanticscholar.org/CorpusID:119743553} {\bibfield  {journal} {\bibinfo  {journal} {The Astrophysical Journal}\ }\textbf {\bibinfo {volume} {153}},\ \bibinfo {pages} {511} (\bibinfo {year} {1968})}\BibitemShut {NoStop}%
\bibitem [{\citenamefont {Chandrasekhar}(1969)}]{CS}%
  \BibitemOpen
  \bibfield  {author} {\bibinfo {author} {\bibfnamefont {S.}~\bibnamefont {Chandrasekhar}},\ }\bibfield  {title} {\bibinfo {title} {Ellipsoidal figures of equilibrium}\ }(\bibinfo {year} {1969})\BibitemShut {NoStop}%
\bibitem [{\citenamefont {{Nduka}}(1971)}]{Nduka}%
  \BibitemOpen
  \bibfield  {author} {\bibinfo {author} {\bibfnamefont {A.}~\bibnamefont {{Nduka}}},\ }\bibfield  {title} {\bibinfo {title} {{The Roche Problem in an Eccentric Orbit}},\ }\href {https://doi.org/10.1086/151194} {\bibfield  {journal} {\bibinfo  {journal} {\apj}\ }\textbf {\bibinfo {volume} {170}},\ \bibinfo {pages} {131} (\bibinfo {year} {1971})}\BibitemShut {NoStop}%
\bibitem [{\citenamefont {Vidal}\ and\ \citenamefont {Cébron}(2017)}]{NewtEllipt}%
  \BibitemOpen
  \bibfield  {author} {\bibinfo {author} {\bibfnamefont {J.}~\bibnamefont {Vidal}}\ and\ \bibinfo {author} {\bibfnamefont {D.}~\bibnamefont {Cébron}},\ }\bibfield  {title} {\bibinfo {title} {Inviscid instabilities in rotating ellipsoids on eccentric kepler orbits},\ }\href {https://doi.org/10.1017/jfm.2017.689} {\bibfield  {journal} {\bibinfo  {journal} {Journal of Fluid Mechanics}\ }\textbf {\bibinfo {volume} {833}},\ \bibinfo {pages} {469–511} (\bibinfo {year} {2017})}\BibitemShut {NoStop}%
\bibitem [{\citenamefont {{Fishbone}}(1972)}]{Rel_RocheI}%
  \BibitemOpen
  \bibfield  {author} {\bibinfo {author} {\bibfnamefont {L.~G.}\ \bibnamefont {{Fishbone}}},\ }\bibfield  {title} {\bibinfo {title} {{The Relativistic Roche Problem}},\ }\href {https://doi.org/10.1086/181006} {\bibfield  {journal} {\bibinfo  {journal} {\apj}\ }\textbf {\bibinfo {volume} {175}},\ \bibinfo {pages} {L155} (\bibinfo {year} {1972})}\BibitemShut {NoStop}%
\bibitem [{\citenamefont {{Fishbone}}(1975)}]{Rel_RocheII}%
  \BibitemOpen
  \bibfield  {author} {\bibinfo {author} {\bibfnamefont {L.~G.}\ \bibnamefont {{Fishbone}}},\ }\bibfield  {title} {\bibinfo {title} {{The relativistic Roche problem. II. Stability theory.}},\ }\href {https://doi.org/10.1086/153349} {\bibfield  {journal} {\bibinfo  {journal} {\apj}\ }\textbf {\bibinfo {volume} {195}},\ \bibinfo {pages} {499} (\bibinfo {year} {1975})}\BibitemShut {NoStop}%
\bibitem [{\citenamefont {Mashhoon}(1975)}]{Mashhoon}%
  \BibitemOpen
  \bibfield  {author} {\bibinfo {author} {\bibfnamefont {B.}~\bibnamefont {Mashhoon}},\ }\bibfield  {title} {\bibinfo {title} {On tidal phenomena in a strong gravitational field},\ }\href {https://api.semanticscholar.org/CorpusID:123096424} {\bibfield  {journal} {\bibinfo  {journal} {The Astrophysical Journal}\ }\textbf {\bibinfo {volume} {197}},\ \bibinfo {pages} {705} (\bibinfo {year} {1975})}\BibitemShut {NoStop}%
\bibitem [{\citenamefont {Shibata}(1996)}]{Sh}%
  \BibitemOpen
  \bibfield  {author} {\bibinfo {author} {\bibfnamefont {M.}~\bibnamefont {Shibata}},\ }\bibfield  {title} {\bibinfo {title} {{Relativistic Roche-Riemann Problems around a Black Hole}},\ }\href {https://doi.org/10.1143/PTP.96.917} {\bibfield  {journal} {\bibinfo  {journal} {Progress of Theoretical Physics}\ }\textbf {\bibinfo {volume} {96}},\ \bibinfo {pages} {917} (\bibinfo {year} {1996})},\ \Eprint {https://arxiv.org/abs/https://academic.oup.com/ptp/article-pdf/96/5/917/5226638/96-5-917.pdf} {https://academic.oup.com/ptp/article-pdf/96/5/917/5226638/96-5-917.pdf} \BibitemShut {NoStop}%
\bibitem [{\citenamefont {Banerjee}\ \emph {et~al.}(2019)\citenamefont {Banerjee}, \citenamefont {Paul}, \citenamefont {Shaikh},\ and\ \citenamefont {Sarkar}}]{TidalNonEq}%
  \BibitemOpen
  \bibfield  {author} {\bibinfo {author} {\bibfnamefont {P.}~\bibnamefont {Banerjee}}, \bibinfo {author} {\bibfnamefont {S.}~\bibnamefont {Paul}}, \bibinfo {author} {\bibfnamefont {R.}~\bibnamefont {Shaikh}},\ and\ \bibinfo {author} {\bibfnamefont {T.}~\bibnamefont {Sarkar}},\ }\bibfield  {title} {\bibinfo {title} {Tidal effects away from the equatorial plane in kerr backgrounds},\ }\href {https://doi.org/https://doi.org/10.1016/j.physletb.2019.05.048} {\bibfield  {journal} {\bibinfo  {journal} {Physics Letters B}\ }\textbf {\bibinfo {volume} {795}},\ \bibinfo {pages} {29} (\bibinfo {year} {2019})}\BibitemShut {NoStop}%
\bibitem [{\citenamefont {Marck}(1983)}]{Ma}%
  \BibitemOpen
  \bibfield  {author} {\bibinfo {author} {\bibfnamefont {J.-A.}\ \bibnamefont {Marck}},\ }\bibfield  {title} {\bibinfo {title} {Solution to the equations of parallel transport in kerr geometry; tidal tensor},\ }\href {http://www.jstor.org/stable/2397341} {\bibfield  {journal} {\bibinfo  {journal} {Proceedings of the Royal Society of London. Series A, Mathematical and Physical Sciences}\ }\textbf {\bibinfo {volume} {385}},\ \bibinfo {pages} {431} (\bibinfo {year} {1983})}\BibitemShut {NoStop}%
\bibitem [{\citenamefont {Carter}(1968)}]{Carter:1968rr}%
  \BibitemOpen
  \bibfield  {author} {\bibinfo {author} {\bibfnamefont {B.}~\bibnamefont {Carter}},\ }\bibfield  {title} {\bibinfo {title} {{Global structure of the Kerr family of gravitational fields}},\ }\href {https://doi.org/10.1103/PhysRev.174.1559} {\bibfield  {journal} {\bibinfo  {journal} {Phys. Rev.}\ }\textbf {\bibinfo {volume} {174}},\ \bibinfo {pages} {1559} (\bibinfo {year} {1968})}\BibitemShut {NoStop}%
\bibitem [{\citenamefont {Bardeen}\ \emph {et~al.}(1972)\citenamefont {Bardeen}, \citenamefont {Press},\ and\ \citenamefont {Teukolsky}}]{Bardeen:1972fi}%
  \BibitemOpen
  \bibfield  {author} {\bibinfo {author} {\bibfnamefont {J.~M.}\ \bibnamefont {Bardeen}}, \bibinfo {author} {\bibfnamefont {W.~H.}\ \bibnamefont {Press}},\ and\ \bibinfo {author} {\bibfnamefont {S.~A.}\ \bibnamefont {Teukolsky}},\ }\bibfield  {title} {\bibinfo {title} {{Rotating black holes: Locally nonrotating frames, energy extraction, and scalar synchrotron radiation}},\ }\href {https://doi.org/10.1086/151796} {\bibfield  {journal} {\bibinfo  {journal} {Astrophys. J.}\ }\textbf {\bibinfo {volume} {178}},\ \bibinfo {pages} {347} (\bibinfo {year} {1972})}\BibitemShut {NoStop}%
\bibitem [{\citenamefont {Wilkins}(1972)}]{OtherSpherical}%
  \BibitemOpen
  \bibfield  {author} {\bibinfo {author} {\bibfnamefont {D.~C.}\ \bibnamefont {Wilkins}},\ }\bibfield  {title} {\bibinfo {title} {Bound geodesics in the kerr metric},\ }\href {https://doi.org/10.1103/PhysRevD.5.814} {\bibfield  {journal} {\bibinfo  {journal} {Physical Review D}\ }\textbf {\bibinfo {volume} {5}},\ \bibinfo {pages} {814 – 822} (\bibinfo {year} {1972})}\BibitemShut {NoStop}%
\bibitem [{\citenamefont {Teo}(2021)}]{Spherical}%
  \BibitemOpen
  \bibfield  {author} {\bibinfo {author} {\bibfnamefont {E.}~\bibnamefont {Teo}},\ }\bibfield  {title} {\bibinfo {title} {Spherical orbits around a kerr black hole},\ }\bibfield  {journal} {\bibinfo  {journal} {General Relativity and Gravitation}\ }\textbf {\bibinfo {volume} {53}},\ \href {https://doi.org/10.1007/s10714-020-02782-z} {10.1007/s10714-020-02782-z} (\bibinfo {year} {2021})\BibitemShut {NoStop}%
\bibitem [{\citenamefont {Ishii}\ \emph {et~al.}(2005)\citenamefont {Ishii}, \citenamefont {Shibata},\ and\ \citenamefont {Mino}}]{Ishii:2005xq}%
  \BibitemOpen
  \bibfield  {author} {\bibinfo {author} {\bibfnamefont {M.}~\bibnamefont {Ishii}}, \bibinfo {author} {\bibfnamefont {M.}~\bibnamefont {Shibata}},\ and\ \bibinfo {author} {\bibfnamefont {Y.}~\bibnamefont {Mino}},\ }\bibfield  {title} {\bibinfo {title} {{Black hole tidal problem in the Fermi normal coordinates}},\ }\href {https://doi.org/10.1103/PhysRevD.71.044017} {\bibfield  {journal} {\bibinfo  {journal} {Phys. Rev. D}\ }\textbf {\bibinfo {volume} {71}},\ \bibinfo {pages} {044017} (\bibinfo {year} {2005})},\ \Eprint {https://arxiv.org/abs/gr-qc/0501084} {arXiv:gr-qc/0501084} \BibitemShut {NoStop}%
\bibitem [{\citenamefont {{Parker}}(1954)}]{1954PhRv...96.1686P}%
  \BibitemOpen
  \bibfield  {author} {\bibinfo {author} {\bibfnamefont {E.~N.}\ \bibnamefont {{Parker}}},\ }\bibfield  {title} {\bibinfo {title} {{Tensor Virial Equations}},\ }\href {https://doi.org/10.1103/PhysRev.96.1686} {\bibfield  {journal} {\bibinfo  {journal} {Physical Review}\ }\textbf {\bibinfo {volume} {96}},\ \bibinfo {pages} {1686} (\bibinfo {year} {1954})}\BibitemShut {NoStop}%
\bibitem [{\citenamefont {{Kochanek}}(1992)}]{noVisc_BNS}%
  \BibitemOpen
  \bibfield  {author} {\bibinfo {author} {\bibfnamefont {C.~S.}\ \bibnamefont {{Kochanek}}},\ }\bibfield  {title} {\bibinfo {title} {{Coalescing Binary Neutron Stars}},\ }\href {https://doi.org/10.1086/171851} {\bibfield  {journal} {\bibinfo  {journal} {\apj}\ }\textbf {\bibinfo {volume} {398}},\ \bibinfo {pages} {234} (\bibinfo {year} {1992})}\BibitemShut {NoStop}%
\bibitem [{\citenamefont {Bildsten}\ and\ \citenamefont {Cutler}(1992)}]{noVisc_BHNS}%
  \BibitemOpen
  \bibfield  {author} {\bibinfo {author} {\bibfnamefont {L.}~\bibnamefont {Bildsten}}\ and\ \bibinfo {author} {\bibfnamefont {C.}~\bibnamefont {Cutler}},\ }\bibfield  {title} {\bibinfo {title} {Tidal interactions of inspiraling compact binaries},\ }\href@noop {} {\bibfield  {journal} {\bibinfo  {journal} {Astrophysical Journal, Part 1 (ISSN 0004-637X), vol. 400, no. 1, p. 175-180.}\ }\textbf {\bibinfo {volume} {400}},\ \bibinfo {pages} {175} (\bibinfo {year} {1992})}\BibitemShut {NoStop}%
\bibitem [{\citenamefont {Shibata}\ \emph {et~al.}(1995)\citenamefont {Shibata}, \citenamefont {Sasaki}, \citenamefont {Tagoshi},\ and\ \citenamefont {Tanaka}}]{Shibata:1994jx}%
  \BibitemOpen
  \bibfield  {author} {\bibinfo {author} {\bibfnamefont {M.}~\bibnamefont {Shibata}}, \bibinfo {author} {\bibfnamefont {M.}~\bibnamefont {Sasaki}}, \bibinfo {author} {\bibfnamefont {H.}~\bibnamefont {Tagoshi}},\ and\ \bibinfo {author} {\bibfnamefont {T.}~\bibnamefont {Tanaka}},\ }\bibfield  {title} {\bibinfo {title} {{Gravitational waves from a particle orbiting around a rotating black hole: PostNewtonian expansion}},\ }\href {https://doi.org/10.1103/PhysRevD.51.1646} {\bibfield  {journal} {\bibinfo  {journal} {Phys. Rev. D}\ }\textbf {\bibinfo {volume} {51}},\ \bibinfo {pages} {1646} (\bibinfo {year} {1995})},\ \Eprint {https://arxiv.org/abs/gr-qc/9409054} {arXiv:gr-qc/9409054} \BibitemShut {NoStop}%
\bibitem [{\citenamefont {Lai}\ \emph {et~al.}(1994)\citenamefont {Lai}, \citenamefont {Rasio},\ and\ \citenamefont {Shapiro}}]{Lai:1993pa}%
  \BibitemOpen
  \bibfield  {author} {\bibinfo {author} {\bibfnamefont {D.}~\bibnamefont {Lai}}, \bibinfo {author} {\bibfnamefont {F.~A.}\ \bibnamefont {Rasio}},\ and\ \bibinfo {author} {\bibfnamefont {S.~L.}\ \bibnamefont {Shapiro}},\ }\bibfield  {title} {\bibinfo {title} {{Hydrodynamic instability and coalescence of binary neutron stars}},\ }\href {https://doi.org/10.1086/173606} {\bibfield  {journal} {\bibinfo  {journal} {Astrophys. J.}\ }\textbf {\bibinfo {volume} {420}},\ \bibinfo {pages} {811} (\bibinfo {year} {1994})},\ \Eprint {https://arxiv.org/abs/astro-ph/9304027} {arXiv:astro-ph/9304027} \BibitemShut {NoStop}%
\bibitem [{\citenamefont {Flanagan}\ and\ \citenamefont {Hinderer}(2008)}]{Flanagan:2007ix}%
  \BibitemOpen
  \bibfield  {author} {\bibinfo {author} {\bibfnamefont {E.~E.}\ \bibnamefont {Flanagan}}\ and\ \bibinfo {author} {\bibfnamefont {T.}~\bibnamefont {Hinderer}},\ }\bibfield  {title} {\bibinfo {title} {{Constraining neutron star tidal Love numbers with gravitational wave detectors}},\ }\href {https://doi.org/10.1103/PhysRevD.77.021502} {\bibfield  {journal} {\bibinfo  {journal} {Phys. Rev. D}\ }\textbf {\bibinfo {volume} {77}},\ \bibinfo {pages} {021502} (\bibinfo {year} {2008})},\ \Eprint {https://arxiv.org/abs/0709.1915} {arXiv:0709.1915 [astro-ph]} \BibitemShut {NoStop}%
\end{thebibliography}%

\end{document}